\documentclass[prd,twocolumn,showpacs,showkeys,preprintnumbers,floatfix,superscriptaddress,longbibliography]{revtex4-1}

\usepackage{amsfonts} 
\usepackage{amssymb} 
\usepackage{amsmath} 
\usepackage{graphicx} 
\usepackage{subfigure} 
\usepackage{array} 
\usepackage{dcolumn} 
\usepackage{bm} 
\usepackage{latexsym} 
\usepackage{longtable} 
\usepackage{hyperref} 
\usepackage{bbold}
\usepackage{mathtools}
\usepackage{soul}

\usepackage[utf8]{inputenc}
\usepackage[braket]{qcircuit}
\usepackage[dvipsnames,usenames]{xcolor}
\usepackage[normalem]{ulem}

\newcommand*{\GF}{G_{\text{F}}}
\newcommand*{\ez}{\vec{e}_{3}}
\newcommand*{\vvec}{\mathbf{v}}
\newcommand*{\I}{\text{i}}
\newcommand*{\dd}{\text{d}}
\newcommand*{\sigv}{\vec{\sigma}}
\newcommand*{\Hvv}{\hat{H}_{\nu\nu}}
\newcommand*{\ax}{\text{ax}}
\newcommand{\swap}[1]{\hat{\rho}_{#1}}
\newcommand\normx[1]{\left\lVert#1\right\rVert}
\newcommand\norm[1]{\big\Vert#1\big\Vert}
\newcommand{\qminner}[2]{\langle #1 \vert #2 \rangle}
\newcommand{\qmmatelem}[3]{\langle #1 \lvert #2 \rvert #3 \rangle}

\newcommand{\UNITN}{{Dipartimento di Fisica, University of Trento, via Sommarive 14, I–38123, Povo, Trento, Italy}}
\newcommand{\TIFPA}{INFN-TIFPA Trento Institute of Fundamental Physics and Applications,  Trento, Italy}
\newcommand{\LANL}{Theoretical Division, Los Alamos National Laboratory, Los Alamos, New Mexico, 87545}
\newcommand{\UNM}{Department of Physics and Astronomy, University of New Mexico, Albuquerque, New Mexico 87131, USA}

\begin{document}
\title{{ Equilibration of quantum many-body fast neutrino flavor oscillations }}
\author{Joshua D. Martin}
\affiliation{\LANL}
\author{Duff Neill}
\affiliation{\LANL}
\author{A. Roggero}
\affiliation{\UNITN}
\affiliation{\TIFPA}
\author{Huaiyu Duan}
\affiliation{\UNM}
\author{J. Carlson}
\affiliation{\LANL}

\preprint{LA-UR-23-28635}
\date{\today}

\begin{abstract}
    Neutrino gases are expected to form in high density astrophysical environments, and accurately modeling their flavor evolution is critical to understanding such environments.  In this work we study a simplified 
    model of such a dense neutrino gas in the regime for which neutrino-neutrino coherent forward scattering is the dominant mechanism contributing to the flavor evolution.
    We show evidence that the generic potential induced by this effect is non-integrable and that the statistics of its energy level spaces are in good agreement with the Wigner surmise.  We also find that individual neutrinos rapidly entangle with all of the others present which results in an equilibration of the flavor content of individual neutrinos.  We show that the average neutrino flavor content can be predicted utilizing a thermodynamic partition function. A random phase approximation to the evolution gives a simple picture of this equilibration. In the case of neutrinos and antineutrinos, processes like $\nu_e  {\bar{\nu}}_e \leftrightarrows \nu_\mu {\bar{\nu}_\mu} $
    yield a rapid equilibrium satisfying $n( \nu_e) n({\bar \nu}_e) = n( \nu_\mu) n({\bar \nu}_\mu) = n( \nu_\tau) n({\bar \nu}_\tau)$ in addition to the standard lepton number conservation in regimes where off-diagonal vacuum oscillations are small compared to $\nu-\nu$ interactions.
\end{abstract}

\maketitle

\section{Introduction}
In hot and dense astrophysical environments such as core collapse supernovae and binary neutron star mergers, neutrinos are 
emitted in such large number fluxes that the average flavor content is important to the dynamic and chemical evolution. Through weak interactions with local nucleons electron flavor neutrinos and antineutrinos can alter the local proton-to-neutron ratio in these environments thereby affecting, for example, $r$-process nucleosynthesis
\cite{Bethe_RevModPhys.62.801,pantaleone1992neutrino,Janka:2006fh,Woosley:2005,Hoffman:1997,Li:2021vqj,Fernandez:2022yyv}.  It is therefore crucial to understand the flavor content of the neutrinos if one wishes to perform detailed studies of the evolution of these systems \cite{Sigl:1992fn,Qian:1994wh,Qian:1995,pastor2002flavor,pastor2002physics,Bell:2003mg,sawyer2004classical,Balantekin:2006tg}.

In regions of high density, neutrinos can undergo coherent forward scattering which is sensitive to the quantum mechanical flavor state of the neutrino.  When scattering on 
charged leptons, a flavor dependent relative phase can develop between components of the flavor state.  The neutrinos can also exchange flavor content with other neutrinos 
through neutral current coherent forward scattering.  When the number density of neutrinos is sufficiently large, this flavor exchange effect is expected to dominate the flavor 
evolution of the neutrino gas, and novel coherent effects such as flavor spectrum splits and swaps may occur \cite{Sigl:1992fn,pantaleone1992neutrino,Qian:1994wh,Qian:1995,pastor2002flavor,pastor2002physics,
Duan:2006jv,Raffelt:2007cb,Raffelt:2007xt}.

In this work, we will study the flavor evolution of such a neutrino gas in the limit that the potential generated by coherent $\nu-\nu$ forward scattering is the 
dominant effect; this is a regime known in the literature as the ``fast" flavor oscillation limit
\cite{sawyer2004classical,PhysRevD.72.045003,PhysRevD.79.105003,PhysRevLett.116.081101,chakraborty2016self,PhysRevLett.118.021101,dasgupta2017fast,PhysRevD.95.103007,PhysRevD.98.043014,PhysRevD.98.103001,airen2018normal,Martin:2019gxb,PhysRevD.101.043009,Roggero2022,PhysRevD.106.103039,Xiong:2021dex}.
In this regime, the redistribution of flavor content among the neutrinos 
is expected to occur on length scales of approximately a meter.  We will study this regime in a quantum many-body formalism with a parametrization of the 
coherent forward scattering potential which does not impose any simplifying symmetries on the momenta of the neutrinos.

In a core-collapse supernovae the neutrino density is governed by $n_{\nu} = L / (  \bar{E} \  4 \pi R^2)$. Taking a total luminosity $L = 10^{53}$ erg/sec, $R = 50$ km,  and an average energy 
$\bar{E} = 10$ MeV gives
a total density $ n_{\nu} \approx 6.6 \times 10^{-7} $ fm$^{-3}$.
For a degenerate relativistic Fermi gas, the average energy is given by $\bar{E} = 3 E_{\textrm{F}} / 4$ where $E_{\textrm{F}}$ is the Fermi energy.  For a two species Fermi gas ($e$ and $\mu$ flavor neutrinos) the number density is given by $n = E_{\textrm{F}}^3 / (3 \pi^{2})$.  Equating the average neutrino energy to the Fermi energy, we find that
the density of a degenerate, zero temperature two-component Fermi gas is approximately $15$ times greater than the above estimated neutrino density.  For 6 neutrino species (3 neutrino and 3 antineutrino species) with roughly equal fractions,
the density is approximately 45
times lower than a degenerate gas of the same average energy.
As we estimate the effects of degeneracy to be minimal, we use Boltzmann statistics for both the neutrinos and antineutrinos.
We will also work in the two-flavor approximation such that 
a single neutrino's flavor quantum state can be written as an SU(2) spinor.

The $\nu -\nu$ coherent forward scattering potential takes the form of an all-to-all Heisenberg-like interaction, and such Hamiltonians are generically expected to be non-integrable except in special cases.  We will provide evidence that this interaction 
is non-integrable in the absence of simplifying symmetries, and that we observe a characteristic signature of (non)integrability when simplifying 
symmetries are imposed (relaxed).  Furthermore, non-integrable Hamiltonians are hypothesized to ``thermalize" in the sense that expectation values of few-body 
operators are expected to equilibrate to a value which can be predicted from an appropriate thermodynamic partition function.  For the generically parameterized potential we observe strong agreement between the one-body flavor expectation values obtained in the exact many-body evolution of the system and those predicted from a grand-canonical Boltzmann distribution.

\section{Energy Spectrum}
The general Hamiltonian governing the coherent evolution of the neutrino flavor content is composed of three parts (see e.g.~\cite{Volpe:2023met}).  The first is the vacuum potential which 
stems from the fact that the neutrino mass states are linear combinations of flavor states.  The second is the matter potential \cite{wolfenstein_PhysRevD.17.2369,mikheev1985resonance} and it is 
generated by coherent forward scattering through charged current interactions with local charged leptons in the environment.  Finally is the $\nu-\nu$ potential generated 
through neutral current coherent forward scattering between neutrinos \cite{Sigl:1992fn,Balantekin:2006tg}. For the rest of this work, we will assume that the vacuum oscillation potential is negligibly small 
relative to the other two potentials, and that its primary effect is to provide perturbation to an otherwise pure flavor product state initial condition.  We also 
assume that the matter profile is uniform in the regions of the environment under consideration, and as such we may consider a co-rotating frame to remove its overall effect on each neutrino.  

After these modifications, only the the $\nu-\nu$ coherent forward scattering Hamiltonian remains.  For $N$ neutrinos it has the form \cite{Balantekin:2006tg}
\begin{equation} \label{eq:Hvv}
    \Hvv = \frac{\mu}{2N} \sum_{i<j}^{N} \left(1 - \vvec_{i} \cdot \vvec_{j} \right) \hat{\sigv}_{i} \cdot \hat{\sigv}_{j} \, .
\end{equation}
Here $\hat{ \sigv }$ is the usual vector of Pauli operators, and $\mu = \sqrt{2} \GF n_{\nu}$ is the scale of the $\nu-\nu$ interaction. Given a typical core-collapse supernovae flux at a radius of ~50 km a simple estimate gives $\mu \approx 1 $ cm$^{-1}$.
As it is the only dimensionful parameter in the problem, we hereafter measure all distances and times in units of $\mu$ and set $\mu = 1$.

We perform simple numerical simulations for small $N$ in periodic boundary
conditions.  This is a toy model because an estimate using 
a typical flux for 20 neutrinos at this density gives a box size of order 300 fm, much too small to be realistic.  This is a standard choice in the literature, 
\cite{Bell:2003mg,sawyer2004classical,Friedland2006,PhysRevD.80.013011,Pehlivan2011,Birol:2018qhx,Patwardhan2019,patwardhan2021spectral,Martin:2021bri,Illa:2022zgu,Xiong:2021evk,shalgar2023evidence,Rrapaj2020,Roggero2021a,Roggero2021b,Roggero2022,Lacroix:2022krq,Cervia2019,PhysRevD.107.043024},
but imposes a great deal more symmetry than a more realistic case.
Relaxing this condition would only induce more decoherence in the simulations, a critical ingredient as we discuss below.
(For more discussion on the role of decoherence see e.g. \cite{Friedland2006,PhysRevD.80.013011,shalgar2023evidence,Johns:2023ewj}.)

We will address two commonly employed simplifying symmetries in the literature.  The first is the uniform coupling symmetry, under which it is assumed that the 
$\vvec_{i} \cdot \vvec_{j}$ term averages to zero under the action of the all-pairs summation in Eq.~\eqref{eq:Hvv}.  For this case, the total $H_{\nu \nu}$ is 
proportional to $J^{2}$ and is straightforwardly diagonalized in the absence of the other two relevant potentials.  For some choices of one-body potentials, the 
uniform couplings Hamiltonian may be diagonalized with an algebraic Bethe ansatz~\cite{Patwardhan2019,Cervia2019}.  This simplification has been well studied so 
we will not consider it further.

Another simplifying assumption is to impose an axial symmetry to the velocity coupling term such that the velocity components orthogonal to the momentum 
symmetry axis (denoted $z$) average to zero.  This is equivalent to making the substitution 
\begin{equation}
    \Hvv \rightarrow \Hvv^{\ax} = \frac{\mu}{2N}\sum_{i<j} (1 - v_{z,i} v_{z,j}) \hat{\sigv}_{i} \cdot \hat{\sigv}_{j} \, .
\end{equation}
It was recently shown that this symmetry imposition leads to an integrable Hamiltonian which can also be diagonalized with an algebraic Bethe ansatz \cite{PhysRevD.107.043024}.

There is no known method for systematically diagonalizing the generic Hamiltonian of Eq.~\eqref{eq:Hvv} through the use of an extensive set of nontrivial conserved charges.  A ``nontrivial'' operator is an operator which is not simply proportional to a projection operator into an eigenstate of the Hamiltonian. 
The only obvious operators which commute with $\Hvv$ are the projections of the total angular momentum
\begin{equation}
    \hat{J}_{\frak{a}} = \frac{1}{2} \sum_{i=1}^{N} \hat{\sigma}_{\frak{a},i}
\end{equation}
(where $\frak{a} \in [1,3]$ is the vector index of the Pauli matrices) and the square of the total angular momentum ($\hat{J}^{2} =\sum_{\frak{a}=1}^{3} \hat{J}_{\frak{a}} \hat{J}_{\frak{a}}$).  
Because $\Hvv$ commutes with $\hat{J}^{2}$, $\hat{J}_{3}$, and the ladder operators $\hat{J}_{\pm}$, the energy eigenstates 
can only be linear combinations of $\ket{j,m}$ in these degenerate subspaces.  
In order to obtain a the full energy spectrum, one can ``brute force" diagonalize $\Hvv$ in the lowest $|m|$ subspace for each $j$.  
Because $\Hvv$ commutes with the angular momentum ladder operators, the energy spectrum is $2j+1$-fold degenerate for each $j$, and eigenstates with higher values of $|m|$ can be systematically constructed from the minimal $|m|$ states with suitable applications of the ladder operators.

In the following, we construct the Hamiltonian by first drawing the $z$ components of the velocities on the 
interval $v_{z} \in \left[-1,1\right]$ 
randomly from a Gaussian-like distribution centered on $v_{z} = 1$ with a standard deviation of $1/2$ such that $P(v_{z}) \propto e^{-2 (v_{z}-1)^{2}}$. We choose this distribution to approximate the forward-peaked-ness of the neutrino momenta emitted from the proto-neutron star surface.  For the axially symmetric case, the $x$ and $y$ components are chosen to be identically zero.  For $\Hvv$, which retains all three components of the unit velocities, we choose $v_{x} = \sqrt{1-v_{z}^{2}} \cos(\phi)$ and $v_{y} = \sqrt{1-v_{z}^{2}} \sin(\phi)$ where $\phi$ is a chosen to be a random number uniformly drawn 
in the interval $\left[0,2\pi\right]$.

Finally, we will solely treat the evolution of initial product states. This necessarily assumes that the neutrinos entering a given region of space have developed a negligible correlation in their flavor evolution prior to the interaction we herein consider.  We make this approximation only in the interest of analyzing the salient dynamics of the evolution under $\Hvv$.  More careful treatment of the initial state will require a detailed study of the neutrino-neutrino self interaction in the presence of non-negligible momentum changing scattering processes in the dense matter of the proto-neutron star core.

\subsection{Level spacings}
We next consider the structure of the spectrum of $\Hvv$ in its $m$ and $j$ symmetry sectors. In each sector we sequentially order the energy eigenvalues and investigate the differences between the sequential energies which we denote $s_{\alpha} \equiv (E_{\alpha+1} - E_{\alpha}) / \bar{s}$ where $\bar{s}$ is chosen such that the average over the energy differences is unity. 

Integrable systems are characterized by an extensive set of (nontrivial) 
operators which commute with the Hamiltonian, and these conserved quantities permit substantial degeneracy in the energy spectrum.  When considering the sequential energy differences for an integrable system, the probability of two sequential energy eigenvalues having spacing $s$ is a Poisson distribution, $P(s) = e^{-s}$.  
In contrast, non-integrable Hamiltonians demonstrate repulsion between energy levels, and the probability distribution for spacing $s$ is, in accordance with the Wigner surmise, of the form $P(s) = \frac{\pi}{2} s e^{- \pi s^2 / 4}$ for a Gaussian orthogonal ensemble (GOE) \cite{scaramazza2016integrable}.

\begin{figure}
    \centering
    \includegraphics[width=0.48\textwidth]{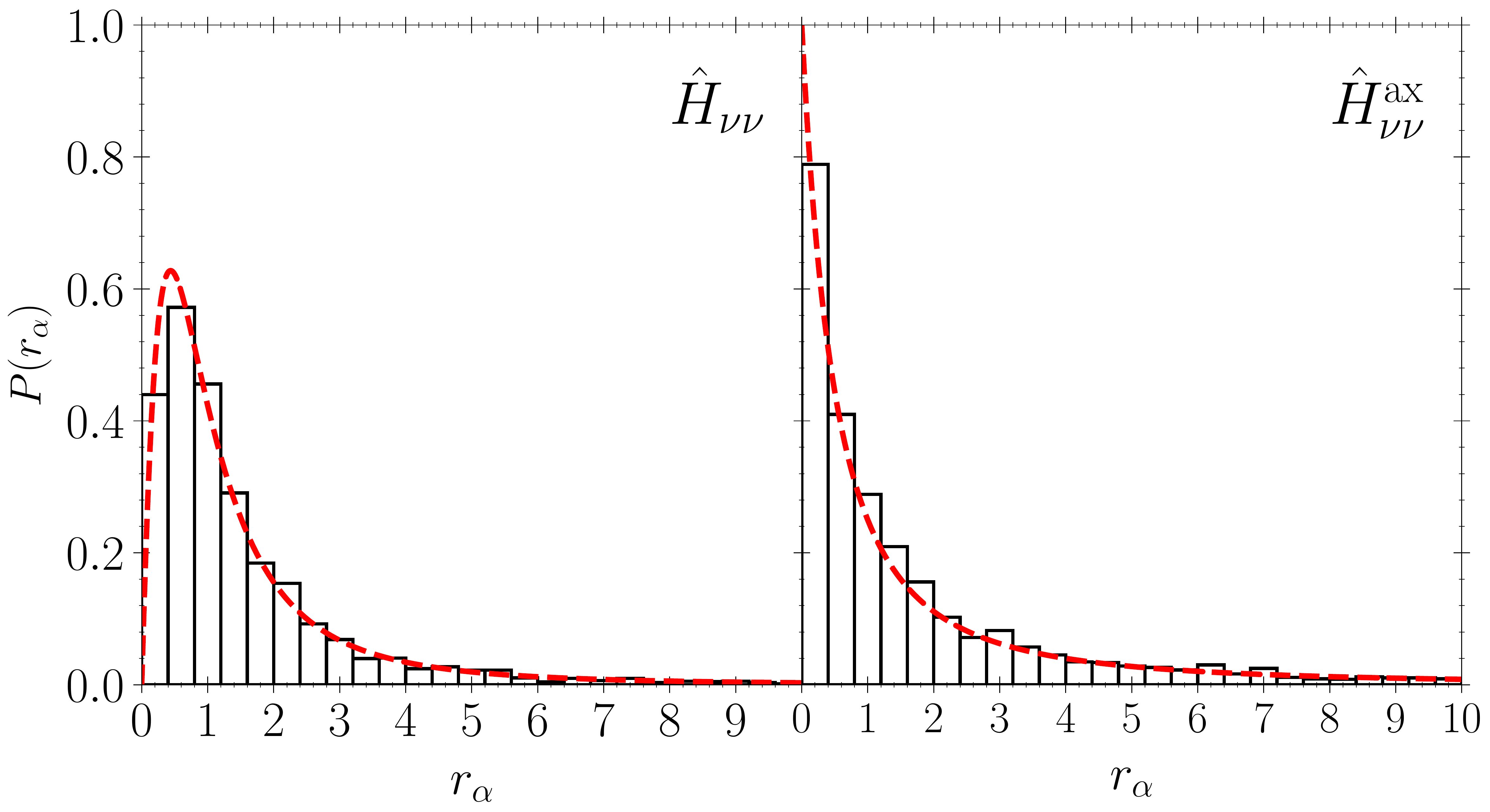}
    \caption{Extracted statistics (black histogram) for the ratio of sequential level spacings of the Hamiltonian 
        for the non-integrable generic parameterization ($\Hvv$, left panel) 
        and for the integrable axially symmetric simplification ($\Hvv^{\ax}$, right panel) 
        for $N = 16$ in the $j = 2$, 
        $m=2$ subspace.  Overlaid as a red dashed line is the predicted probability distribution 
        for the ratio of sequential level spacings for the non-integrable (GOE, left panel) distribution and the integrable (Poisson, right panel) distribution.}
    \label{fig:levelStats}
\end{figure}

When extracting the distribution of energy differences from a particular Hamiltonian, an unfolding procedure must be performed in order to approximately 
remove the effect arising from the fact that the Hamiltonian has a finite density of states which may be sparser in some energy regions than others \cite{atas2013distribution}. 
Instead of considering directly the distribution of level differences $s_{i}$, one can consider the probability distribution for the ratio of sequential 
level differences, defined as $r_{\alpha} \equiv \frac{s_{\alpha+1}}{s_{\alpha}}$ which is insensitive to the local density of states. The new probability distribution, $P(r)$, 
for the integrable (Poisson) case is
\begin{equation} \label{eq:p_r_Poisson}
    P(r) = \frac{1}{(1+r)^{2}} \, .
\end{equation}
While for non-integrable (GOE) cases, $P(r)$ is
\begin{equation} \label{eq:p_r_GOE}
    P(r) \approx \frac{27}{8} \frac{r(1+r)}{(1 + r(1+r))^{\frac{5}{2}}} \, .
\end{equation}
For the GOE case, the expression in Eq.~\eqref{eq:p_r_GOE} was derived as an exact expression for $3 \times 3$ random matrices drawn from a Gaussian Orthogonal Ensemble.  It is in good agreement with numerical studies of larger random matrices, and the correction derived from fits to numerical distributions extracted from larger random matrices can be found in \cite{atas2013distribution}.  The approximate correction for the large matrix case is $\mathcal{O}(10^{-2})$ and is not visible on the scale plotted in Fig.~\ref{fig:levelStats}.

In Fig.~\ref{fig:levelStats} we compare the distribution of $r_{i}$ for the two Hamiltonians we have discussed.  The Hamiltonian, $\Hvv$, implements generic unit magnitude three velocities, while the axially symmetric (abbreviated ``$\ax$") Hamiltonian, $\Hvv^{\ax}$, implements the identical $v_{z,i}$ values as $\Hvv$, but takes the $x$ and $y$ components of the velocities equal to zero by construction.  As previously mentioned, $\Hvv^{\ax}$ is known to be integrable, and therefore the level-spacings in its spectrum are expected to obey Poisson-like statistics.
For both cases, we use 16 neutrinos in the construction of the Hilbert space.  

We diagonalize $\Hvv$ and $\Hvv^{\ax}$ in the $j=2$ subspace and plot the normalized histogram of ratios of level spacings ($r_{\alpha}$).  We also compare directly with the expected universal probability distributions for the integrable and non-integrable cases.  This figure provides evidence that the generically parameterized Hamiltonian does not have an extensive set of ``hidden" conserved charges corresponding to some hitherto unrecognized symmetry.

To validate that the level spacing statistics in the other $j$ subspaces behave similarly, we compute the average $\bar{r}_{\alpha}$ with respect to the extracted probability distribution and show the results for $j \leq 4$ in table ~\ref{tab:r_avg}. The distribution of ratios of level spacings should not depend on the subspace, so this average value should take a universal value.  For the GOE case, it can be shown that the value is $\bar{r}_\alpha \approx 1.7789$ (see table 1 of \cite{atas2013distribution}).  However, for the Poisson case the average diverges as $r_\alpha \rightarrow \infty$, so for any finite distribution of level spacings the average can take arbitrary and unbounded values.

\begin{table}[thb]
    \caption{Average ratios of consecutive level spacings for $\Hvv$ and $\Hvv^\ax$ of 16 neutrinos in the subspaces with $j=0,\ldots,4$ and $m=0$ which have 1430, 3432, 3640, 2548, and 1260 energy levels, respectively.}\label{tab:r_avg}
    \begin{ruledtabular}
    \begin{tabular}{c | c  c  c  c  c } 
        $j$ & 0 & 1 & 2 & 3 & 4 \\
        \hline 
        $\bar{r}(\Hvv)$ & 1.6651 & 1.7126 & 1.7512 & 1.7707 & 1.8508 \\
        $\bar{r}(\Hvv^\ax)$ & 3.7211 & 3.7741 & 3.8224 & 3.6091 & 3.8764 
    \end{tabular} 
    \end{ruledtabular}
\end{table}

\section{Energy Dephasing}
The distribution of the ratio of level spacings is in good agreement with the probability distribution one would expect if the Hamiltonian under consideration was a random matrix drawn from a GOE.  The level repulsion resulting from a lack of an extensive set of symmetries implies that energy differences should not be expected to vanish for the bulk of states in the spectrum. 
Furthermore the effective random nature of the Hamiltonian implies that differences in energies should not generically be simple rational numbers.  As such, the time-dependent phases of off-diagonal matrix elements of operators in the energy basis should be expected to individually average to zero over arbitrary time windows after the state has evolved for a sufficiently long time.  Thus, under a time average over an arbitrary interval at late time, the expectation value of $\hat{\sigma}_{3,i}$ should be in good agreement with the value predicted by computing the average with respect to the incoherent energy diagonal distribution which preserves the energy state probabilities of the initial condition.  Thus we will compute and compare the approximation
\begin{align} \label{eq:micro_avg}
    &\frac{1}{\Delta} \int_{t_{\text{eq}}}^{t_{\text{eq}} + \Delta} \dd t \langle \hat{\sigma}_{3,i} \rangle (t) = \nonumber \\
        &\frac{1}{\Delta} \int_{t_{\text{eq}}}^{t_{\text{eq}} + \Delta} \dd t \sum_{E,E'} e^{-\I (E-E')t} \qminner{E}{\psi_{0}} \qminner{\psi_{0}}{E'} \qmmatelem{E'}{\hat{\sigma}_{3,i}}{E} \nonumber \\
        &\approx \sum_{E} |\qminner{E}{\psi_0}|^{2} \qmmatelem{E}{\hat{\sigma}_{3,i}}{E} 
\end{align}
where $\ket{\psi_{0}}$ represents the initial quantum state and $t_{\text{eq}} > 1 / \mu$ is the (as yet unknown) time it takes for the system to 
reach its equilibrium value. The time window 
$ \left[ t_{\text{eq}}, t_{\text{eq}} + \Delta \right] $ 
must also be selected such that the oscillations due to the coherent evolution of the quantum system about the average expectation value can be fully captured.  We will refer the value $t_{\text{eq}}$ as the equilibration timescale, and we will discuss our observations of it in the next section.  We refer to the incoherent probability weighted sum of operator expectation values in the final line of Eq.~\eqref{eq:micro_avg} as the energy mixed state distribution (EMSD).

\subsection{Thermalization}
Non-integrable many-body Hamiltonians are widely expected to obey the so-called ``Eigenstate Thermalization Hypothesis'' (ETH)~\cite{PhysRevA.43.2046,Srednicki1994,Srednicki_1996}, see~\cite{d2016quantum} for a review. Many numerical studies of ETH have been conducted, we refer to Refs. \cite{2009PhRvA..80e3607R,2010PhRvA..82a1604R,2014PhRvE..90e2105K} as a very incomplete list of examples for when the system has an underlying lattice structure. For systems with all-to-all interactions or no discernible lattice structure, there is a less extensive literature, and we refer to Refs. \cite{2017PhRvL.118l7201B,2021PhRvB.104u4203L,2022Entrp..25....8V} as examples.
Besides integrable systems, for a generic Hamiltonian, ETH may only hold for the majority of the eigenstates. Certain eigenstates (quantum scars) embedded in the spectrum may have additional symmetries beyond that of the Hamiltonian, invalidating ETH for that state, see for instance~\cite{shiraishi2017systematic,pakrouski2020many}. However, these states are expected to have measure zero in the density of states in the thermodynamic limit, unless there is special symmetry protection. Loading the Hamiltonian with additional symmetry, while not inducing integrability, results in the fracturing of the Hilbert space into many block-diagonal subsectors, with each subspace being simply too small for ETH~\cite{regnault2022quantum}.

The ETH is fundamentally a statement about the structure of ``few-body'' operators that act on a sub-extensive number of the local Hilbert spaces that tensor together to form the full many-body Hilbert space. For a ``few-body'' operator $\hat{O}$, the matrix element between two eigenstates $|E_{\alpha}\rangle,|E_{\beta}\rangle$ of the Hamiltonian is 
hypothesized to be 
 given by the ansatz:
\begin{align}
\label{eq:eth_matel}
\langle E_{\alpha}|\hat{O}|E_{\beta}\rangle&= \hat{O}(E_{\rm ave})\delta_{\alpha\beta}\nonumber\\
&\qquad+e^{-S(E_{\rm ave})/2}f_{\hat{O}}(E_{\rm ave},\omega_{\alpha\beta})R_{\alpha\beta}\,,\\
S(E_{\rm ave})&=\text{ln}\sum_{\alpha}E_{\alpha}\delta_{\epsilon}(E_{\rm ave}-E_{\alpha})\,.
\end{align}
where $E_{\rm ave}=\frac{E_{\alpha}+E_{\beta}}{2}$, and $\omega_{\alpha\beta}=E_{\alpha}-E_{\beta}$. $\hat{O}(E_{\rm ave})$ is the microcanonical average of the operator $\hat{O}$ over states near $E_{\rm ave}$, and $S$ is the suitable microcanonical entropy, counting the number of eigenstates near $E_{\rm ave}$ within a small window given by $\epsilon$. $R_{\alpha\beta}$ is an independent random number for each $\alpha,\beta$ with average zero, but variance $1$. When the dimension of the total Hilbert space is large, we can expect $S$ to be large even for very narrow windows, so that in the thermodynamic limit, we can take $\epsilon\rightarrow 0$ and have the matrix element of the operator in eigenstates be dominated by the microcanonical average. Finally, the function $f_{\hat{O}}$ is related to the linear response correlation function of the thermodynamic system when perturbed by the operator $\hat{O}$~\cite{srednicki1999approach} and, for fixed $E_{\rm ave}$, decreases exponentially in the energy difference $\omega_{\alpha\beta}$ for such chaotic systems (see e.g.~\cite{2019PhRvX9d1017P,PhysRevLett.123.230606}). 

\begin{figure}
    \centering
    \includegraphics[width=0.48\textwidth]{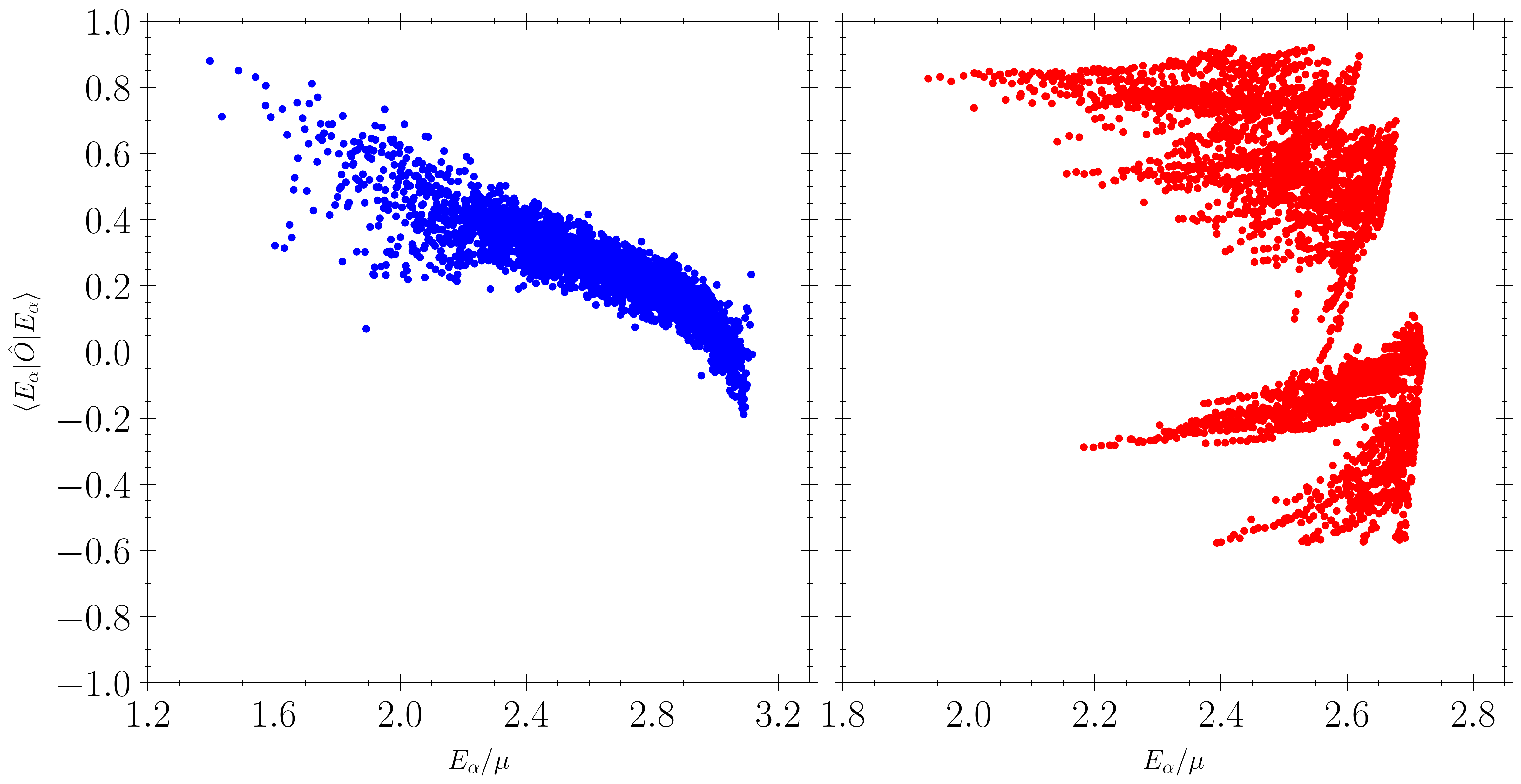}
    \caption{Diagonal matrix elements $\langle E_\alpha\lvert \hat{O}\rvert E_\alpha\rangle$ with $\hat{O}=\hat{\sigma}_{3}$ on the first neutrino in the energy eigenbasis as a function of the energy $E_\alpha$ (in units of $\mu$). The left panel shows results for the non-integrable Hamiltonian $\Hvv$ while the right panel shows results for the integrable case $\Hvv^{\ax}$, in both cases we only look at the $j=2$ and $m=2$ subspace. }
    \label{fig:eth_matel_diag}
\end{figure}

\begin{figure}[b]
    \centering
    \includegraphics[width=0.48\textwidth]{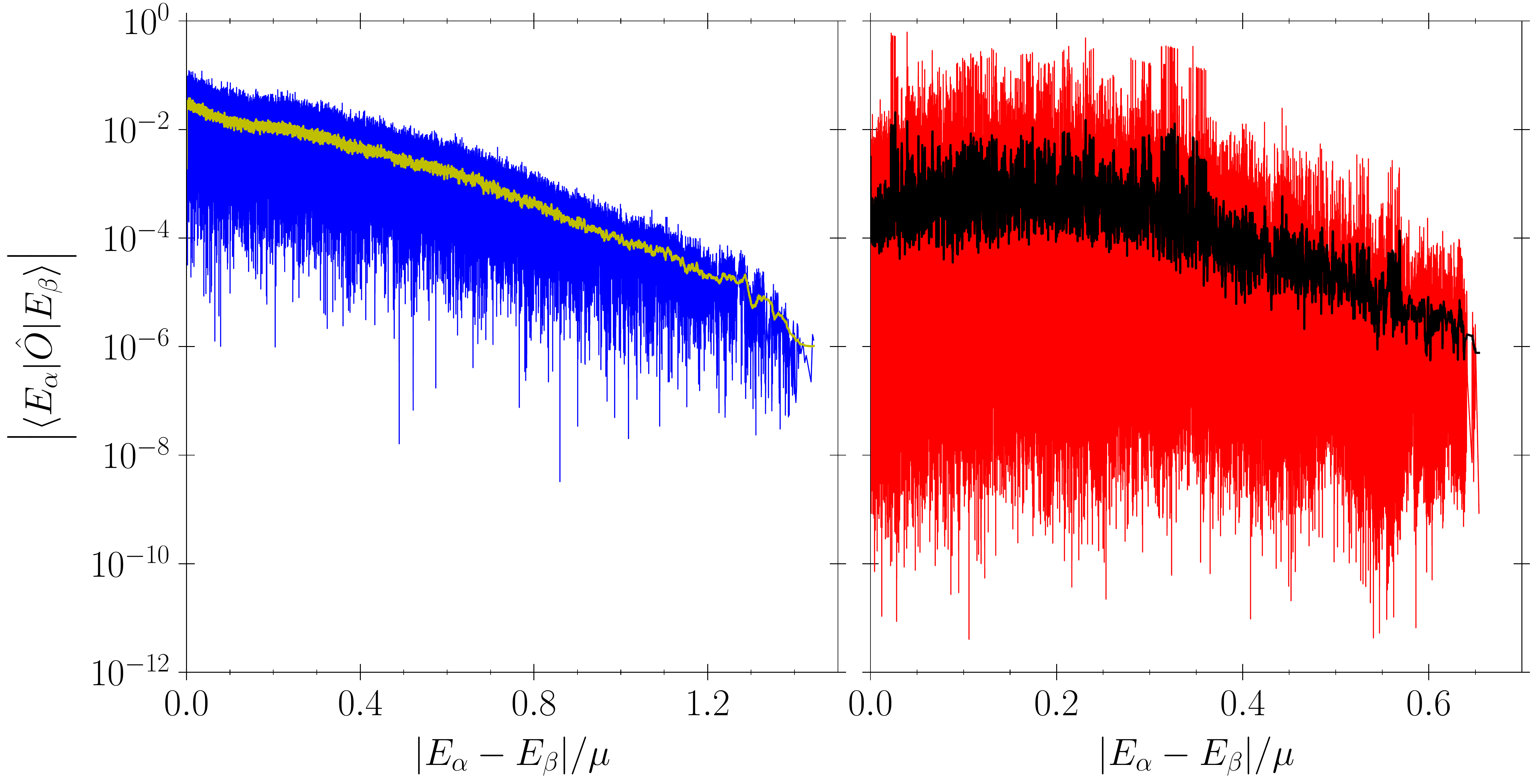}
    \caption{Magnitude of the off-diagonal matrix elements $\left|\langle E_\alpha\lvert \hat{O}\rvert E_\beta\rangle\right|$ with $\hat{O}=\hat{\sigma}_3$ on the first neutrino in the energy eigenbasis as a function of the absolute value of the energy difference $\left|E_\alpha-E_\beta\right|$ (in units of $\mu$). The left panel shows results for the non-integrable Hamiltonian $\Hvv$ (blue data) while the right panel shows results for the integrable case 
    $\Hvv^{\ax}$ (red data), in both cases we only look at the $j=2$ and $m=2$ subspace. Also shown is the running average of the matrix elements' magnitude obtained taking the median over a window of $50$ matrix elements (yellow curve on the left panel and black curve on the right panel).}
    \label{fig:eth_matel_odiag}
\end{figure}

In general we expect the ETH ansatz (Eq.~\eqref{eq:eth_matel}) for the matrix elements to hold for a chaotic system but not for an integrable one. As shown above, the neutrino Hamiltonian $\hat{H}_{\nu\nu}$ in Eq.~\eqref{eq:Hvv} we use can reproduce both behaviors depending on the choice of the velocities. To test the validity of ETH for our problem we consider matrix elements of $\hat{\sigma}_{3,i}$ on the first neutrino in both the integrable and non-integrable regime. We first show the diagonal matrix elements $\langle E_\alpha\lvert \hat{O}\rvert E_\alpha\rangle$ in Fig.~\ref{fig:eth_matel_diag} for both the integrable (red data on right panel) and non integrable case (blue data on left panel). One can see that in the former (integrable) case the diagonal matrix element has large fluctations over nearby frequencies throughout the whole spectrum while for the non-integrable case the expectation value in the bulk of the spectrum is approximately a function of energy alone as expected from the ansatz in Eq.~\eqref{eq:eth_matel}. Next, in Fig.~\ref{fig:eth_matel_odiag} we show the magnitude of the off-diagonal matrix elements of $\hat{\sigma}_{3,i}$ on the first neutrino as a function of the energy difference $\lvert\omega_{\alpha\beta}\rvert$ for all states in a narrow energy window of size $\epsilon=0.005[\mu]$ around an average energy $E_{\rm ave}$ in the middle of the spectrum: for the results shown here we took $E_{\rm ave}=2.4[\mu]$. One can clearly see an exponential decay for large energy difference in the non-integrable case (left panel) while for the integrable model (right panel) the size of the off-diagonal matrix elements fluctuates by more than six orders of magnitude for every value of $\lvert\omega_{\alpha\beta}\rvert$. To better visualize the importance of a large number of outliers in the integrable case, which are instead not present in the non-integrable system on the left, we also report in Fig.~\ref{fig:eth_matel_odiag} the running average of the matrix element size obtained by taking the median over a window of $50$ matrix elements. This is a direct estimate of the absolute value of the function $f_{\hat{O}}(E,\omega)$ in Eq.~\eqref{eq:eth_matel} for $E$ in the middle of the spectrum (cf.~\cite{d2016quantum,PhysRevLett.111.050403}).

All results shown in Fig.~\ref{fig:eth_matel_diag} and Fig.~\ref{fig:eth_matel_odiag} were obtained for a system with $N=16$ neutrinos and for states restricted to $j=2$ and $m=2$ subspace.

The basic intuition behind eigenstate thermalization is that few-body operators are unable to distinguish which eigenstate the system is in, with nearby (in energy) eigenstates displaying similar few-body behavior. When the dimension of the Hilbert space becomes large, the eigenstate thermalization is often described using canonical or grand-canonical ensembles, since the expectation values are expected to be effectively equal in either ensemble. When using the (grand)-canonical ensemble, the temperature (or other chemical potentials) are tuned to match the energy and other quantum numbers of the desired eigenstate of the system. After this, the matrix elements of generic few-body operators can be computed in the appropriate thermal ensembles. When ETH holds, we can effectively claim that up to small corrections, the time-averaged spectrum of few-body operators ``thermalize'' even in a single eigenstate, and are thus computable from thermal ensembles. One can generalize the eigenstate thermalization to any state of the system, provided expectation value of the variance of the energy for that state is appropriately small.

If our system ``thermalizes" in this sense, then we can predict the (time-averaged) expectation value of the individual neutrinos at large times utilizing a grand-canonical statistical distribution.  As the system is time independent, and $\hat{J}_{3}$ and $\hat{J}^{2}$ commute with the Hamiltonian, we construct a partition function using one temperature parameter ($\beta$) and two chemical potentials ($\mu_{3}$ and $\mu_{2}$).  For a given initial condition, we find these parameters by fitting the expectation values of the relevant operators calculated using the partition function to the invariant expectation values calculated with respect to the initial condition.  This amounts to finding $\beta$, $\mu_{3}$, and $\mu_{2}$ such that
\begin{equation} \label{eq:fitSystem}
    \bra{\psi_{0}} \hat{O} \ket{\psi_{0}} = \frac{1}{\mathcal{Z}} \text{Tr} \left( \hat{O} e^{- \beta \Hvv + \mu_{3} \hat{J}_{3} + \mu_{2} \hat{J}^{2} } \right)
\end{equation}
for the conserved quantities $\hat{O}=\Hvv, \hat{J}_{3}, \hat{J}^2$ and
$\mathcal{Z} = \text{Tr} \left(e^{- \beta \Hvv + \mu_{3} \hat{J}_{3} + \mu_{2} \hat{J}^{2} } \right)$ is the partition function. Once the temperature and chemical potentials have been determined, the (time-averaged) expectation value of the flavor content for a given neutrino ($i$) can be predicted by computing $\langle \hat{\sigma}_{3,i} \rangle \approx \text{Tr}\left(\hat{\sigma}_{3,i} e^{-\beta \Hvv + \mu_{3} \hat{J}_{3} + \mu_{2} \hat{J}^{2}}\right)$.

We next investigate the late time behavior of an example product state initial condition for 16 neutrinos.  We employ the same couplings in $\Hvv$ as in the previous sections, and we order them from lowest to highest value of $v_{z,i}$ in increasing values of the particle index $i \in [1,16]$.  We next choose the first $i \in [1,\lfloor N/2 \rfloor+1]$ neutrinos to be electron flavor (corresponding to the lowest 9 $v_{z}$ values for 16 neutrinos), and the remaining (highest $v_{z}$) neutrinos to be $\tau$ flavor.  Finally, to mimic the effects of the neglected one body contributions to the Hamiltonian, we perturb the initial flavor configuration by a small random polar rotation away from initial pure flavor states, as well as a random rotation in the $x-y$ plane of the flavor Bloch sphere.  The first rotation mimics the effect of the small effective mixing angle induced by non-commutation of the dense matter and vacuum potentials, while the second rotation mimics the phase accumulation from the rapid rotations about the flavor axis induced by the matter potential.

\begin{figure}
    \centering
    \includegraphics[scale=0.33]{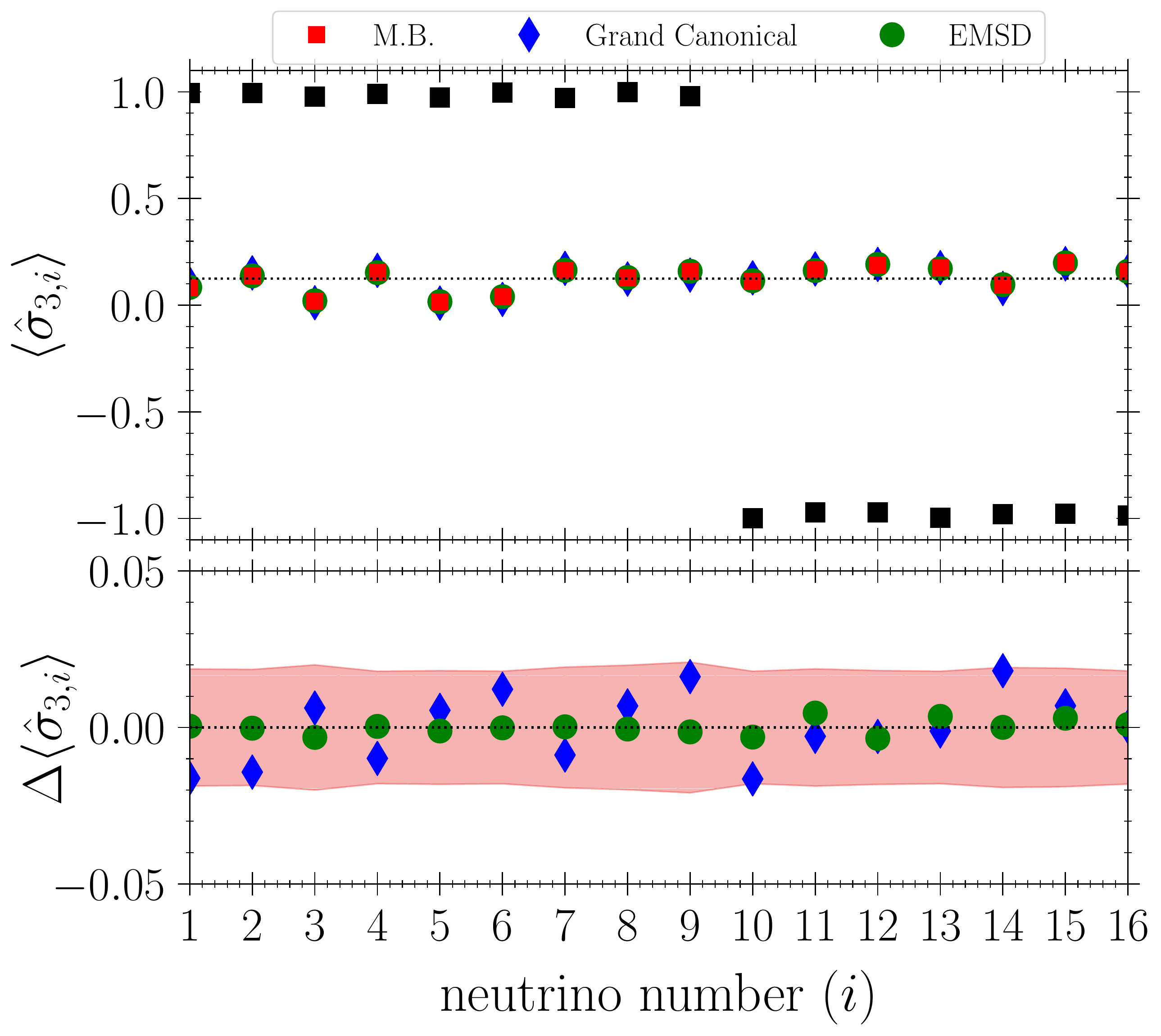}
    \caption{In the top panel we show the initial condition (black squares) and the average over the late time oscillations of $\langle \hat{\sigma}_{3,i} \rangle$ obtained from the solution to the many-body Schr\"{o}dinger equation for each neutrino. The thin black dotted line indicates the conserved value $2 \langle \hat{J}_{3} \rangle / N$. Blue diamonds are the expectation value predicted from the grand-canonical partition function fit from the initial condition, while green circles indicate the energy mixed state distribution (EMSD) average. 
    The lower panel shows the difference ($\Delta \langle \hat{\sigma}_{3,i} \rangle$) between the exact solution and the partition function fit (blue diamond) and the numerical solution to the MB Schr\"{o}dinger equation and the microcanonical approximation (green circles).  The filled red area represents the RMS deviation of the oscillations of the exact solution about its mean.}
    \label{fig:sigz_th}
\end{figure}

Once the couplings and initial conditions are specified, we then evolve the 16 neutrino quantum state by numerically solving the many-body Schr\"{o}dinger equation in the interval 
$t = \left[0,10^{3}\right]\mu^{-1}$. Using the time evolved quantum state, we compute the average of $\langle \hat{\sigma}_{3,i} \rangle$ over the interval $t = [10^{1},10^{3}] \mu^{-1}$.  We show this time averaged quantity in the top panel of Fig.~\ref{fig:sigz_th} and we compare time averaged numerical predictions for $\langle \hat{\sigma}_{3,i} \rangle$ with the mixed state predictions for each spin. The red squares represent the time-averaged values of the evolved quantum state, and the green circles (blue diamonds) show the expectation value estimated using the 
EMSD (grand-canonical) approximation of Eq.~\eqref{eq:micro_avg} (Eq.~\eqref{eq:fitSystem}). The bottom panel shows the difference between the late time averaged many-body solution and the two energy diagonal predictions, while the filled red region indicates the root-mean-square (RMS) deviation of the time oscillations of the expectation value of the many-body solution about its mean (as seen in Fig. \ref{fig:sigz_v_t}). 
We observe that both of the energy diagonal approximations for the estimation of $\langle \hat{\sigma}_{3,i} \rangle$ are fully within the RMS oscillations of the the many-body solution about its mean, excepting one outlying point estimated using the grand-canonical ensemble.

Its important to note that while the interaction is $SU(2)$ invariant, we choose a particular spin direction along which to quantize (which is informed by the initial polarization) and denote it as $\ez$.  Because $\Hvv$ commutes with 
$\hat{J}_{+}$ and $\hat{J}_{-}$ their expectation values are also conserved, but these operators cannot be simultaneously diagonalized with $\hat{J}_{3}$ and $\hat{J}^{2}$.  Therefore no mixed state which is \textit{energy diagonal} will be able to accurately predict their expectation values.  For the initial conditions we consider here, 
$\langle \hat{J}_{+} \rangle \approx \langle \hat{J}_{-} \rangle \approx 0$.  
As such, the energy-diagonal mixed states we consider here are adequate for describing the polarization configurations.  However in the presence of substantial orthogonal polarization (i.e. substantially nonzero values of $\langle \hat{J}_{\pm} \rangle$), a more careful treatment of the non-abelian conserved charges would be necessary \cite{Murthy:2022dao}.

\subsection{Approach to equilibrium}
In the previous sections we have argued that in the absence of simplifying symmetries the $\nu-\nu$ interaction Hamiltonian in its various symmetry sectors has a level spacing distribution which is in good agreement with that of a 
random Hamiltonian drawn from the GOE.  We have also argued that this implies generic product state initial conditions should display substantial dephasing in energy such that one-body expectation values 
can be computed from an energy-diagonal distribution.  Once dephased, the one-body expectation values are expected to achieve an equilibrium 
value from which they stray only transiently and with small amplitude. 

\begin{figure}
    \centering
    \includegraphics[scale=0.35]{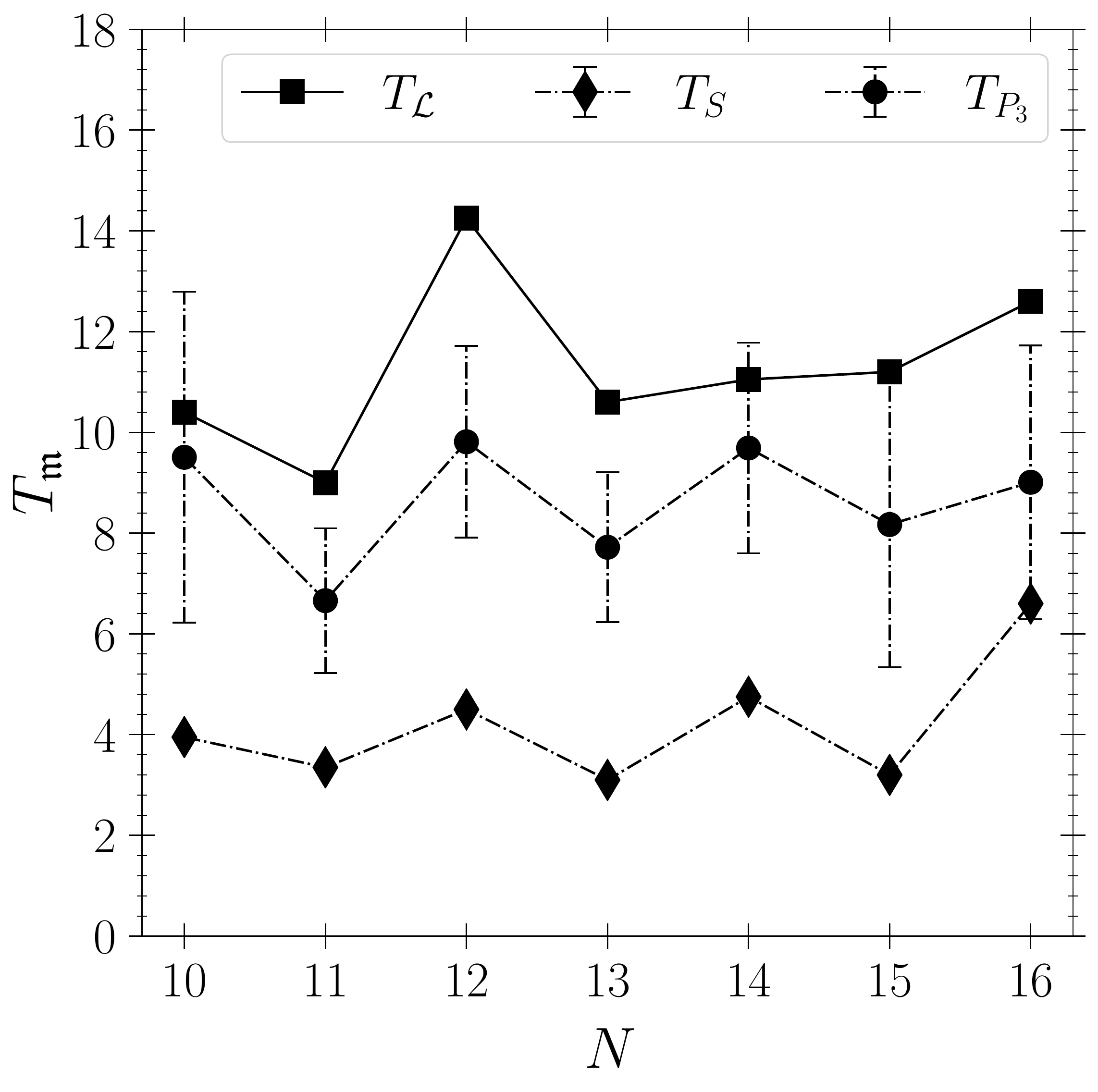}
    \caption{Times for which: the average time ($T_{S}$, indicated by black diamonds) the one-body entropies reach $95$\% of their maximum; the Loschmidt echo ($T_{\mathcal{L}}$, black squares) reaches its first local minimum; and the average time ($T_{P_{3}}$, indicated by black circles) the the one-body $\langle \hat{\sigma}_{3}\rangle$ reaches its first turning point. }
    \label{fig:Nscaling}
\end{figure}

To investigate the approach to this equilibrium, we evolved seven different systems each with $N$ spins from $N=10$ to $N=16$ in integer steps.  For each system size we independently selected velocities and initial conditions in the same manner as described for the $N=16$ case shown in detail above.  We then considered three simple measures of the speed of the evolution.
First, for each system size ($N$), we consider the one-body von Neumann entropy ($S_{i} = -\text{Tr}(\rho_{i} \log_{2}(\rho_{i}))$) of each neutrino, and we find the earliest time for which $S_{i} > 0.95$.  We then take the average of these times over all the spins for a given system size, we denote this average large entropy time as $T_{S}$, and plot $T_{S}$ as
black diamonds in Fig.~\ref{fig:Nscaling}.   We also computed the standard deviation of $T_{S}$ over the spins for each system size, however in each case the error bar would be too small to resolve on the plotted scale of Fig.~\ref{fig:Nscaling}.

\begin{figure}
    \centering
    \includegraphics[scale=0.35]{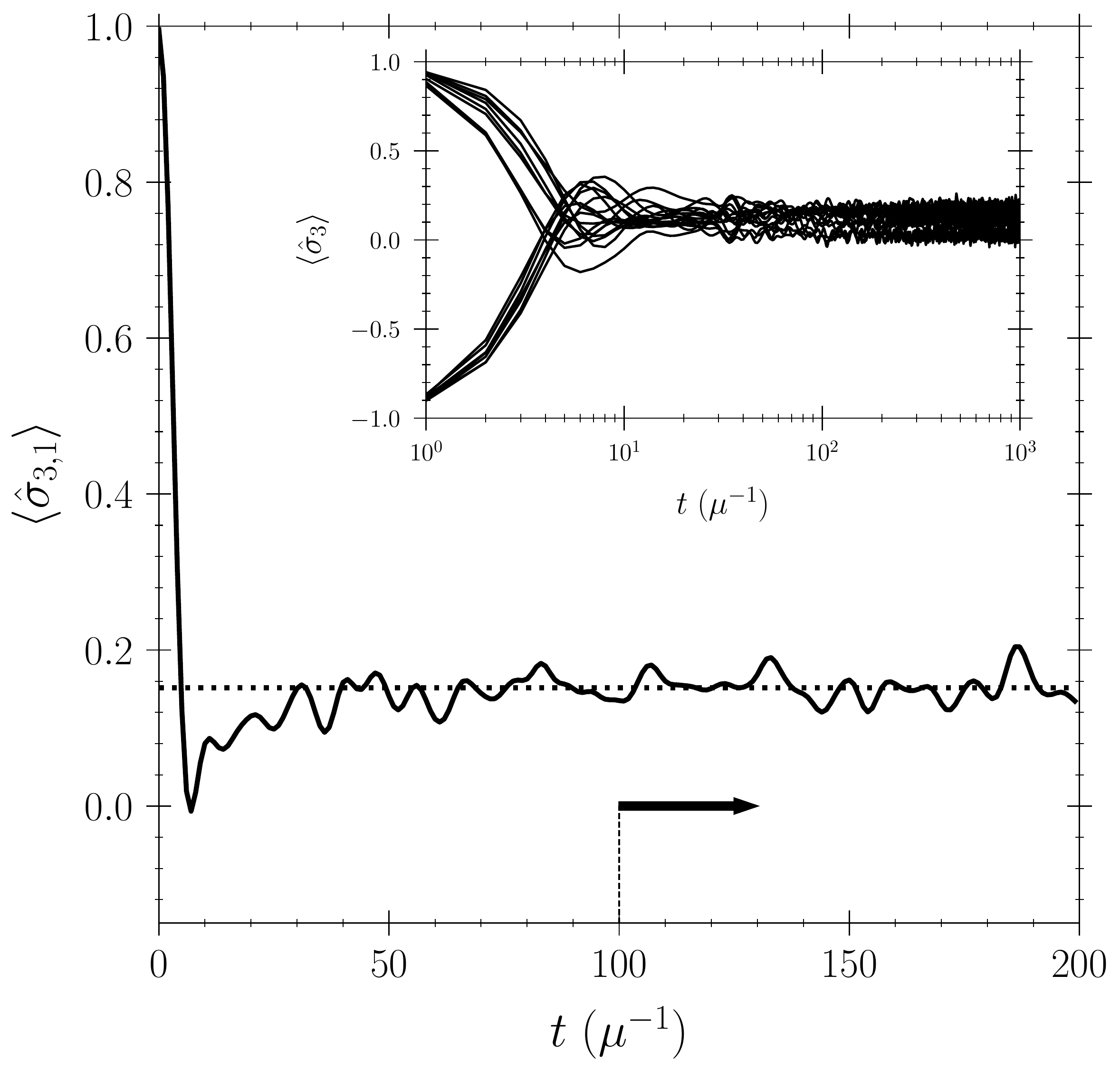}
    \caption{Evolution of $\langle \hat{\sigma}_{3} \rangle$ for neutrino $i=1$ compared to the prediction utilizing the grand-canonical partition function (black dotted line) at early time.  The line and arrow indicate when the time-averaging begins for the evaluation of the data plotted in Fig.~\ref{fig:sigz_th}.
    \textit{Inset:} $\langle \hat{\sigma}_{3} \rangle$ as a function of time for every neutrino on the entire considered time domain. }
    \label{fig:sigz_v_t}
\end{figure}

Next, we note the time dependent behavior Fig.~\ref{fig:sigz_v_t} of $P_{3} \equiv \langle \hat{\sigma}_{3} \rangle$.  After  evolving away from the initial value, the expectation value crosses the thermal prediction before turning and approaching it again. We observe this behavior across all of the chosen system sizes and for each spin.  We therefore consider, for each system size, the time at which the value $\langle \hat{\sigma}_{3} \rangle$ for each spin reaches a turning point at which its first time derivative vanishes.  We then average these turning-point times over all the spins, denote the average time as $T_{P_{3}}$, and plot it as black circles in Fig.~\ref{fig:Nscaling}, and we indicate the standard deviation from the average value over the spins in a given system size with error bars. 

Finally, we consider the Loschmidt echo $\mathcal{L}(t) = |\langle\psi_0\lvert e^{-\I \Hvv t}\rvert\psi_0\rangle|^{2}$ which quantifies the probability of measuring the time evolved state in the initial configuration.  The $t=0$ curvature of the Loschmidt echo is given by the variance of the Hamiltonian, and the earliest time at which it (the Loschmidt echo) can vanish is bounded by the quantum speed limit~\cite{Giovannetti_2003b}.  For chaotic systems, the Loschmidt echo is expected to saturate to an equilibrium value which scales inversely proportionally to the size of the system's Hilbert space~\cite{PhysRevE.79.046211}, a scaling behavior that we have verified for $\Hvv$. The dynamics of the echo at intermediate times is an active area of study, but we investigate the time of the first minima in the echo as a proxy to the equilibration timescale of the system.

For all of the calculations we have performed, we have observed an approach to an equilibrium value in the one-body flavor content with a timescale which appears insensitive to the total number of spins. 
The equilibration we observe in flavor content is subsequent to a development of one-body entanglement on a timescale which is similarly insensitive to $N$.  Our numerical observations are consistent 
with a time-to-equilibrium which scales simply as $\mathcal{O}(\mu^{-1})$.
In Fig.~\ref{fig:Nscaling} we show the timescales we extracted for the equilibration of these quantities for a range of system sizes.  Because our computational method retains all amplitudes, we are computationally limited in the system sizes $N$ we can investigate. 

For states that are close to polarized product states in the $\ez$ direction, we can argue that the time-scale to equilibrium cannot scale to zero as $N\gg 1$. 
In the Appendix, we calculate to quadratic order in the Taylor series expansion of the time evolution of $\langle\hat{\sigma}_{3,i}(t)\rangle$ about $t = 0$. We find that, for the typical states we consider here, the linear term in the expansion vanishes, and the quadratic term scales as $\mu^2 / N$.
We expect such a Taylor series to have a radius of convergence scaling as $\mu^{-1}$. Noting $\langle \hat{\sigma}_{3,i}(t=0) \rangle\approx\pm 1$, then truncating the series at quadratic order and estimating when $\langle \hat{\sigma}_{3,i}(t) \rangle \approx 2 \langle \hat{J}_{3} \rangle / N\ll1$, we would conclude $t\sim \sqrt{N}/\mu$. However, such a time-scale is outside the expected region of convergence for our series, making such a conclusion not self-consistent. We can, though, conclude that $t\rightarrow 0$ as $N\gg 1$ is impossible: the smallness of the quadratic term, and the suppression of higher orders as $t\rightarrow 0$, would not allow a self-consistent solution to the equilibrium condition in arbitrarily small neighborhoods of the origin. We intend to follow up these observations with investigations of these timescales both analytically and with computational methods which may allow substantially larger system sizes, such as tensor network methods.

\subsection{Path integral description of  time evolution and equilibrium distributions}

In the results presented in the previous section, it is striking that at late times
each individual neutrino flavor expectation value is approximately given by
$\langle \hat{\sigma}_{3,i} \rangle \approx 2 \langle \hat{J}_{3} \rangle / N$.
We observe this approximate flavor isotropy even when there is substantial correlation between the initial 
flavor content and momenta of the neutrinos (as in the initial split configuration of Fig.~\ref{fig:sigz_th}).  This is behavior we have observed across a variety of system sizes, and initial correlations between flavor and momenta. 
The total $\langle  \hat{J}_3 \rangle $ is of course conserved by the Hamiltonian,
but there is no a priori reason to expect all the spins to reach the same equilibrium value.
To clarify this point and to define the equilibrium values for general systems of neutrinos 
(and both neutrinos and anti-neutrinos), we consider the time evolution in a path integral approach.

The time-evolved expectation value of an operator expressed as an expansion of overlaps with the initial state
for a system described by the GOE produces random phases in the off-diagonal elements
that would normally be present in Eq. \eqref{eq:micro_avg}. Rewriting the expectation value as a sum over product
states of $\nu_e$ and $\nu_{\tau}$ spins 
$ \ket{ n }$  the amplitude of the state in state $\ket{ n }$ as a
function of time is:
\begin{equation}
\qminner{n}{ \psi (t) } =  \sum_\alpha e^{ -\I E_\alpha t } \qminner{n}{ E_\alpha } 
    \qminner { E_\alpha }{ \psi_0 }
\end{equation}
where we have taken the initial state $\ket{\psi_0}$ as an arbitrary state in the $\ket{n}$ basis.

Equivalently, writing the state $ \ket{ \psi (t)  }$ as a sum over the states $\ket{n}$
\begin{equation}
\ket{\psi (t) } = \sum_n A_n(t) e^{ \I  \phi_n (t) } \ket{ n },
\end{equation}
where the $A_n (t) $ are the magnitudes of the overlaps (real and positive) and the phases are $\phi_n (t) $ .
The magnitudes and phases are given by:
\begin{equation}
A_n (t) e^{ \I \phi_n (t) } = \sum_\alpha e^{ - \I E_\alpha t} \qminner{ n }{ E_\alpha } 
    \qminner{ E_\alpha }{ \psi_0 }.
\end{equation}

The expectation value of $\hat{\sigma}_{3,i}$ averaged over a time window of size $\Delta$ is:
\begin{align}
\frac{1}{\Delta} &\int_{t}^{t+\Delta} \dd t \langle \psi (t ) \vert \hat{\sigma}_{3,i} \vert \psi(t) 
    \rangle \approx  \nonumber \\
&\sum_{n,\alpha} 
\bra{n} \hat{\sigma}_{3,i} \ket{n} 
\vert \qminner { n }{E_\alpha} \qminner{ E_\alpha }{ \psi_0 } \vert^2 ,
\end{align}
and we have used the facts that $\hat{\sigma}_{3,i}$ is diagonal in the $\ket{n}$ basis and that the average
over non-diagonal energy eigenstates over time goes to zero.
This is a sum over diagonal matrix elements in the $\ket{n}$ basis, each with a positive coefficient, suggesting
the time-evolved state can be written as an incoherent sum of the states $\ket{n}$. 

For an initial state with overlaps with many eigenstates, the random phases ($e^{ -\I E_\alpha t} $) for a GOE would
translate to random phases in the $\nu_{e}-\nu_{\tau}$ product state basis  $\phi_n (t)$. 
This would not be true near the ground state where the phases cannot be random, but initial product states we wish to describe are not near the ground state of this Hamiltonian. Random phases also keep $\langle \hat{\sigma}_{\pm,i}  \rangle = \langle \hat{\sigma}_{\mp,i} \rangle = 0$
for all times for an initial product state in $\ket{n}$.

The Hamiltonian can always be divided into diagonal ($\Hvv^{\textrm{d}}$) and off-diagonal 
($\Hvv^{\textrm{od}}$)
pieces, where, in the $\ket{n}$ basis, the off-diagonal pieces are proportional to $(1 - \vvec_i \cdot \vvec_j  ) \hat{\sigv}_i \cdot \hat{\sigv}_j $ which is simply a weighted permutation operator exchanging anti-parallel flavor spins $i$ and $j$.
The full path integral describing the propagation
with $\exp [ - \I \Hvv t] $ is then a sum over all paths with all-to-all two-body spin exchanges while the diagonal piece just reduces to pure phases, as do two-body exchanges between parallel spins.

We can write the path integral as a sum over
all powers of $\Hvv^{\textrm{od}}$ operators.
Summing over the number of non-diagonal operators at random times, with diagonal evolution (pure
phases) between the non-diagonal terms, we can rewrite the time evolution with a path integral as

\begin{widetext}

\begin{align} \label{eq:path_int}
\langle n | \exp[-\I \Hvv t] | \psi_0 \rangle &=
  \sum_m \frac{(-\I )^m } {m!} \sum_{n_i \cdots n_m} \int \dd t_0 \cdots \dd t_m \nonumber 
    \langle n \vert \exp[-\I \Hvv^{\textrm{d}} t_m] \Hvv^{\textrm{od}}  
    \vert n_m \rangle  \\ &  \langle n_{m} \vert \exp[ - \I \Hvv^{\textrm{d}} t_{m-1} ] \Hvv^{\textrm{od}} \vert n_{m-1} \rangle \cdots
\langle n_{1} \vert \exp [ -\I \Hvv^{\textrm{d}} t_{1} ] \Hvv^{\textrm{od}} \vert n_0 \rangle 
    \langle n_0 \vert \psi_0 \rangle.
\end{align}

\end{widetext}

Here, we can see that $\Hvv^{\textrm{d}}$ introduces pure phases into each term
of the path integral, while the off-diagonal terms $ \Hvv^{\textrm{od}}$
induce transitions from one product state to another.
In this equation, the sum of all times $t_i$ has a resultant of $t$, and
we have separated the path integral into
terms with a specific
number of insertions (indexed by $m$) of the off-diagonal operators.

In principle one could sample these paths by their absolute magnitudes as is done in many quantum Monte Carlo (QMC) approaches, in particular diagrammatic Monte Carlo \cite{VanHoucke:2010}. If the initial product state connects to many basis states and the
$\Hvv^{\textrm{d}}$ introduces random phases, the final state is
an incoherent sum of product states. The off-diagonal terms
$\Hvv^{\textrm{od}}$ induce transitions as a series of spin swaps between one product state and another.
Such a sampling of swaps conserves the expectation value of $\hat{J}_{\pm}, \hat{J}_{3}$ and $\hat{J}^2$ of an initial product
state since exchanges of the original spins have the same expectation values of total and projected spin as the original state.

All permutations of the original spins can be reached by a series of two-body
swaps. For an initial product state of many neutrinos, we
can easily calculate the expectation value of 
$\langle \Hvv^k \rangle$ for small $k$. These quantities
must be conserved by the time evolution. 
If the angular distributions are similar, as they
are deep inside a proto-neutron star, all product states
of permutations of the initial spins should, when
averaged over time, have equal magnitudes of overlaps $A$.
Each permutation of the initial product state will
have approximately the same $\langle \Hvv^{\textrm{d}} \rangle$, with a variance
inversely proportional to the number of neutrinos since
there are $N^2$ terms in the Hamiltonian. 
Unitarity requires that the average of the squared
absolute magnitudes is inversely proportional to the
number of spins.

More generally the time-averaged magnitudes of the amplitudes
are expected to depend on the $ \langle \Hvv^{\textrm{d}} \rangle $
of the individual permutations. These can arise, for example, from
interference between different insertions of off diagonal operators leading
to the same final state.  If we assume time-averaged overlaps vary smoothly
with $\langle \Hvv^{\textrm{d}} \rangle$, we can calculate this dependence by requiring that the
lowest $N$ moments of the Hamiltonian are conserved. The linear dependence could
be parametrized as a temperature as in ETH, because we also know that the absolute
magnitudes of the time-averaged amplitudes are real and positive.  Higher order 
moments put further constraints on the evolution. Example calculations
for small numbers of neutrinos are discussed below. Crucially the low order moments of
$\Hvv$ can be calculated exactly from the initial product state, or from an incoherent sum of
physically reasonable product states.

The path integral of Eq.~\eqref{eq:path_int} can be described as 
a random walk in the basis states. For large times we expect the
phases of individual  $\nu_e-\nu_{\tau}$ product states to become
random if the initial product state is in the middle of the spectrum.
The short time evolution of an amplitude of a particular state ($a_n(t) \equiv \qminner{n}{\psi(t)}$) up to order $\delta t$ is governed by
\begin{align}
a_n (t + \delta t) = \exp[-\I \Hvv^{\textrm{d}} \delta t] a_n(t) + \sum_{m \in P(n)} \bra{n} \Hvv^{\textrm{od}} \ket{m} \delta t .
\end{align}
There are $\mathcal{O}(N^2)$ terms in the sum over states $m \in P(n)$ where $P(n)$ is the set all of states which can be produced by one pair permutation in $\Hvv^{\textrm{od}}$, each matrix element with a substantial random
component.

Assuming phases of individual $\nu_e-\nu_{\tau}$ product states at large enough time
separations are random, the expression for the path
integral can be described as a random walk in the
basis states.  The complex amplitude for a specific
state at a given
large time will be distributed as a two-dimensional
Gaussian centered at zero describing its real and imaginary parts.
The average magnitude (squared) is governed by unitarity.
Integrating over times after equilibration should
produce a constant absolute magnitude of the overlap,
with a variance decreasing approximately inversely proportional
to the square root of the time integrated over.
Expectation values are obtained as the incoherent sum of expectation values
in individual purely $\nu_{e}$-$\nu_\tau$ product states.

\begin{figure}
\centering
\includegraphics[scale=0.33]{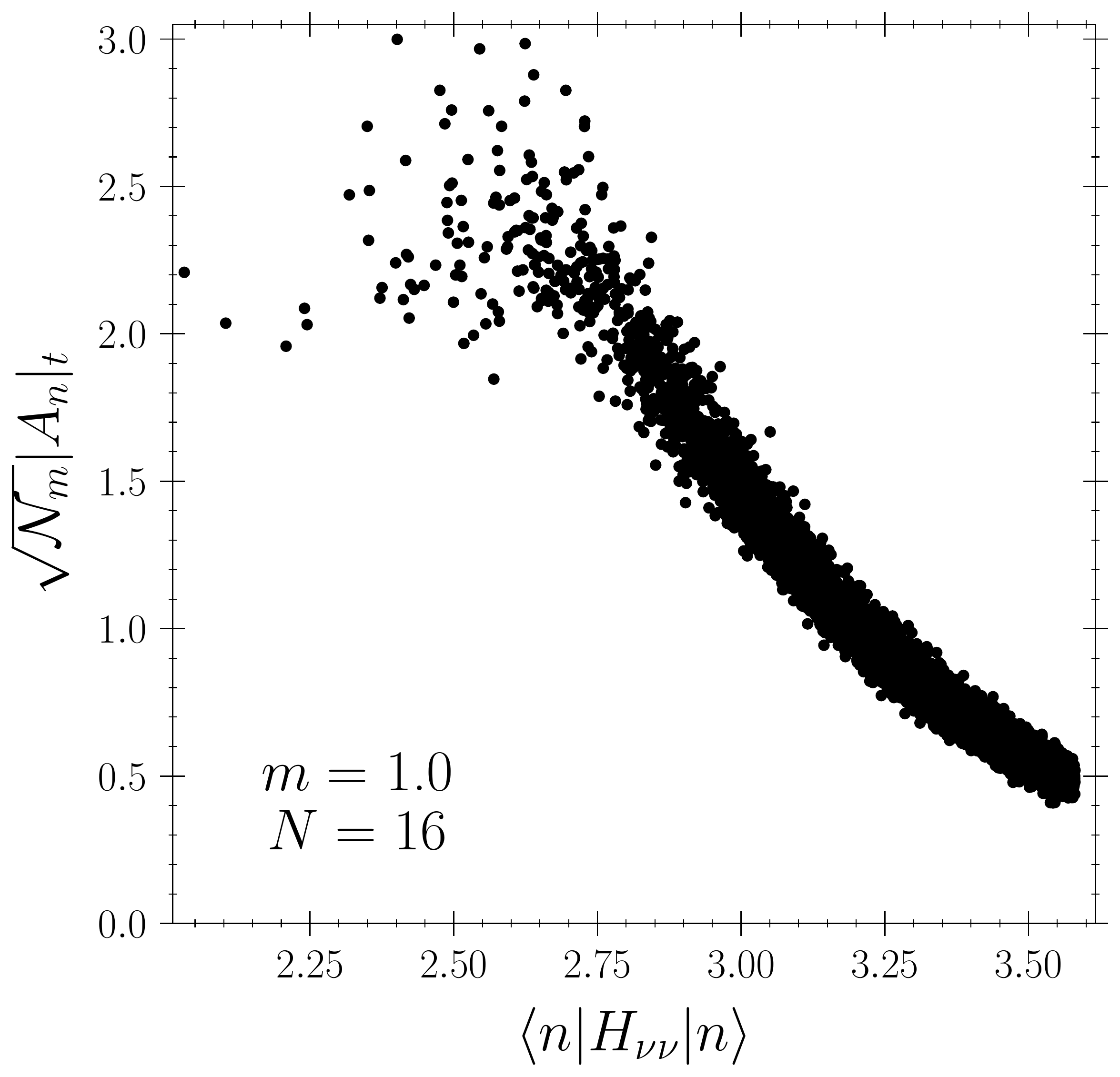} \\
\includegraphics[scale=0.33]{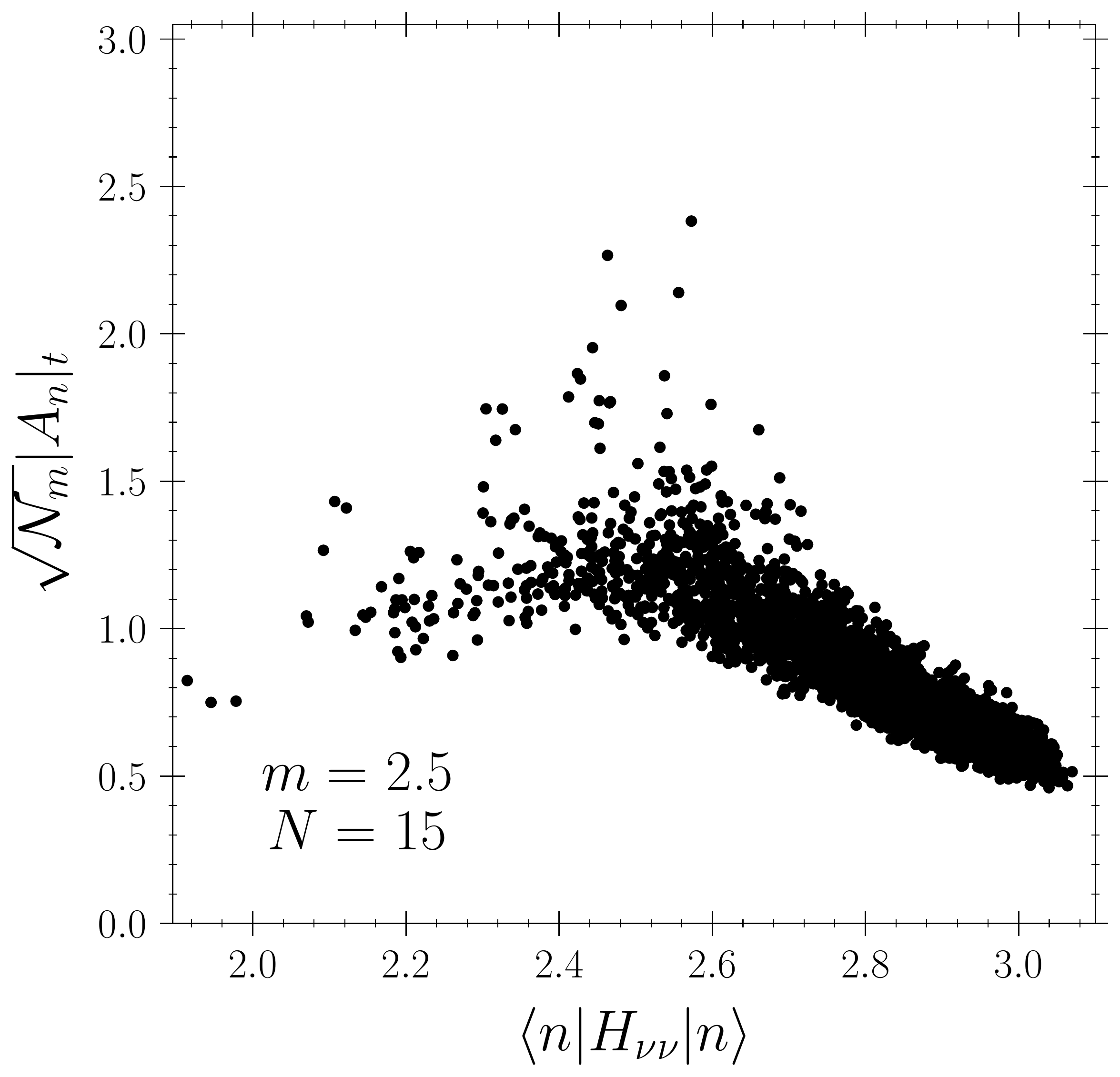}
\caption{Time-averaged magnitudes of state amplitudes ($|A_{n}|_{t}$) for the $N=16$ (top) and $N=15$ (bottom) cases in the $m$ subspaces (top: $m=1$, and bottom: $m=2.5$) with the largest total overlap with the total state plotted against the respective diagonal elements of the Hamiltonian in the product state ($\ket{n}$) basis. In each case, the state projected into the specified $m$ subspace has been normalized to unity, and the time-averaged amplitude magnitudes have been scaled by $\sqrt{\mathcal{N}_{m}}$ where $\mathcal{N}_{m}$ is the total number of states in the specified $m$ subspace.} \label{fig:psiA_HnDiag}
\end{figure}

We investigated the behavior of the time average of $\vert A_{n}(t) \vert$ at late times exactly for small $N$.  The above discussion suggests that we should expect $\vert A_n \vert_{t} \propto \frac{1}{\sqrt{\mathcal{N}_{m}}}$ where the $t$ subscript indicates a time average, and $\mathcal{N}_{m}$ is the total number of states in the Hilbert space with quantum number $m = \bra{n} \hat{J}_{3} \ket{n}$.  In Fig.~\ref{fig:psiA_HnDiag} we show two cases.  The top panel represents $\sqrt{\mathcal{N}_{m}} \times \vert A_{n} \vert_{t}$ for the quantum state whose one-body expectation values are shown in Figs. \ref{fig:sigz_th} and \ref{fig:sigz_v_t} for all states $\ket{n}$ in the $m=1$ subspace, plotted against the corresponding diagonal elements of $\Hvv$.  The lower panel is similar but for an $N=15$ state which initially had purely $10$ $\nu_{e}$ and $5$ $\nu_{\tau}$ resulting in unit overlap with the $m=2.5$ subspace. For the lower panel, spins were chosen $\nu_{e}$ and $\nu_{\tau}$ randomly, the $v_{x/y}$ components of velocity were chosen with random azimuthal angles on the unit sphere, and the $v_{z}$ components were chosen randomly from the uniform interval $\left[0,1 \right]$.

This figure shows that for both cases, in the respective $m$ subspaces, the time-averaged magnitudes $\vert A_{n} \vert_t \approx \mathcal{O}(1) / \sqrt{\mathcal{N}_{m}}$.  Both of the considered cases show some structure in the magnitudes versus $\bra{n} \Hvv \ket{n}$, as they must in order to conserve all of the moments of the Hamiltonian. 
For a more complex
Hamiltonian with terms violating the conservation of $J^2$ or components of $J$ or for time-varying Hamiltonians  the symmetries are further reduced
\cite{Martin:2023ljq}, making the approximations considered here even more accurate.
Even for the case of a specific product initial state and a static
Hamiltonian with these symmetries the behavior of the absolute magnitudes $A$ is smooth in energy.

The equilibrium distribution in energy and angle reached by a pure neutrino system is simply the
weighted average of the initial distributions in energy and angle.  For the simple two-body
interaction discussed here, the total number of neutrinos of each flavor is conserved.
This equilibration would produce the horizontal dashed lines in Fig.~\ref{fig:sigz_th} and is reached quite quickly, as we discussed previously. 

For an incoherent sum over orthogonal states, the path integral
can be modeled as a classical process including exchanges of
pairs of spins.  A classical swap network can be implemented
for a large number of neutrinos.  The potential dependence
of the amplitudes  on $\langle \Hvv \rangle$ could be implemented
through a Metropolis Monte Carlo with the parametrization of the
energy dependence determining the accept/reject probability.
For cases with minimal dependence, such as equal initial
angular distributions, the accept/reject probability could be
fixed to 1. 

Because of the lack of coherence, it could be that non-forward scattering would also be important. The magnitude of these amplitudes would be similar to the forward scattering. However the phases from different magnitudes of the neutrinos' momenta
would oscillate rapidly on the MeV$^{-1}$ scale, rendering their contributions
very small. In principle all amplitudes should be consistently included and summed but these highly 
oscillatory pieces are unlikely to significantly interfere with the forward
scattering amplitudes. The same picture of random diagonal phases and rapid flavor exchanges would occur
and  not change this picture, particularly since the angular dependence is not measurable. At larger distances where the neutrino flux is reduced, off-diagonal vacuum oscillation terms will be important, leading to further flavor evolution. Potentially this could be modeled as a system of single neutrino
spins evolving according to their one-body Hamiltonian with random swaps imposed. However the equilibrium we discuss should dominate at smaller radii where much of the dynamics and chemical evolution occurs. At larger radii the evolution is governed by simple one-body dynamics due to the expected geometric reduction in the strength of the two-body interaction.

The case of a mixture of neutrinos and antineutrinos of different flavors is more interesting.
The total number density of a particular flavor of neutrino (or antineutrino) is denoted by $n(\nu_{x})$, and the net lepton 
flavor is given by $n(\nu_{x}) - n(\bar{\nu}_{x})$.
The net lepton number is preserved by the path integral picture with random phases and two-particle exchanges.
The conservation of these lepton number conservation conditions (totaling three, one for each flavor) plus total number conservation yield only four conditions, though, when six
must be satisfied to define equilibrium, neutrinos and antineutrinos of three flavors each. 
Assuming random phases renders the quantum problem essentially classical, then it is possible to determine the equilibrium distributions very easily.

Assume that the flavor swapping processes $\nu_e(\mathbf{p}) + \nu_x(\mathbf{p}') \rightleftharpoons \nu_e(\mathbf{p}') + \nu_x(\mathbf{p})$ have achieved equilibrium. Because the forward and reverse processes have the same scattering amplitude, one must have $n_{\nu_e}(\mathbf{p}) n_{\nu_x}(\mathbf{p}') = n_{\nu_e}(\mathbf{p}') n_{\nu_x}(\mathbf{p})$ or, equivalently, $n_{\nu_e}(\mathbf{p}) / n_{\nu_x}(\mathbf{p}) = n_{\nu_e}(\mathbf{p}') / n_{\nu_x}(\mathbf{p}') = q$. 
Therefore, the $\nu_e$ to $\nu_x$ ratio is independent of the neutrino momentum when the flavor equilibrium is obtained. Similarly, one also has $n_{\nu_e}(\mathbf{p}) / n_{\nu_x}(\mathbf{p}) = n_{\bar\nu_x}(\mathbf{p}') / n_{\bar\nu_e}(\mathbf{p}') = q$ if $\nu_e(\mathbf{p}) + \bar\nu_e(\mathbf{p}') \rightleftharpoons \nu_x(\mathbf{p}) + \bar\nu_x(\mathbf{p}')$ have reached equilibrium. This is consistent with the flavor isospin notation where the antineutrinos are treated as neutrinos with negative energies \cite{Duan:2005cp}. Using the constancy of $q$ one can determine the equilibrium distribution completely for given $n(\nu_{e}) - n({\bar{\nu}}_{e})$, and $n(\nu_{x}) - n({\bar{\nu}}_{x})$.

In environments where both neutrinos and anti-neutrinos are present, 
flavor off-diagonal evolution from
processes like
$\nu_e  {\bar{\nu}}_e \leftrightarrows \nu_\mu {\bar{\nu}_\mu} $
yields a rapid equilibrium. In such an equilibrium state, the time averaged
flux into a particular state must equal the time averaged flux out of a state.
Since the magnitude of the Hamiltonian matrix elements are symmetric under time reversal this implies the probability
of the product of the densities of neutrinos and anti-neutrinos in each flavor
must  satisfy $n( \nu_e) n({\bar \nu}_e) = n( \nu_\mu) n({\bar \nu}_\mu) = n( \nu_\tau) n({\bar \nu}_\tau)$, where $n$ represents the density of neutrinos of a given flavor. These two additional conditions on the product of neutrino and antineutrino densities,
along with the four previous, completely determine the equilibrium condition.

\begin{figure}
    \centering
    \includegraphics[width=0.48\textwidth]{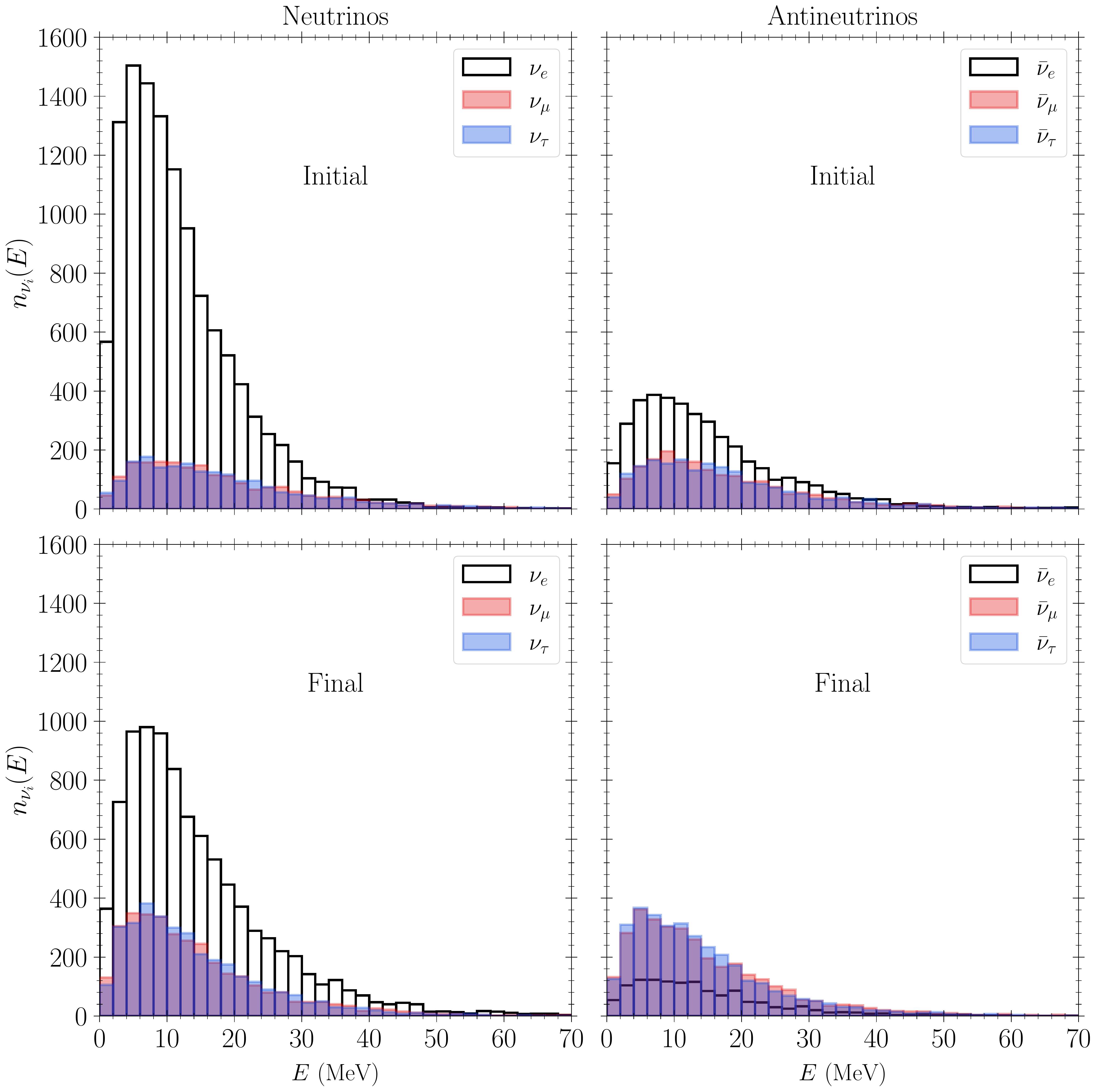}
    \caption{Initial (top) and Final (bottom) neutrino energy spectra $(n_{\nu_{i}} (E))$ for each flavor (denoted i) in a simple calculation assuming random phases from the diagonal parts of the Hamiltonian. The product of the number of electron neutrinos and antineutrinos equilibrates with the other flavors, while the difference is maintained for each flavor.}
    \label{fig:pairswaps}
\end{figure}

We consider a simply implemented three-neutrino-flavor swap network of 24000 total neutrinos and antineutrinos, each labeled only by their energy, as we expect that the flavor evolution 
will rapidly tend to isotropy in momentum given the above arguments.  We distribute them 
randomly in energy with a distribution that mimics an energy weighted Boltzmann distribution 
of the form $\mathcal{P}(E) \propto E e^{- \beta E}$ which we show in the top panels for all 
neutrino and antineutrino species
in Fig. \ref{fig:pairswaps}. We start with initial distributions 
with fractional $n(\nu_e)= 1/2$ and
an average energy of 11.9 MeV, $n(\bar{\nu}_e) = 1/6$ with an average energy of 15 MeV, and all other species with $n(\nu_{x}) = 1/12$ and an average energy of $18.2$ MeV.  

With these initial populations, we perform swaps among the neutrinos and antineutrinos of different energies by randomly selecting two neutrinos from the distribution and swapping their energies.  In this process, if a neutrino and antineutrino of the same type are selected, we also allow them to change to a different flavor, with probability $\frac{1}{3}$ for each flavor.
We perform this swapping process on average 250 times per neutrino in the distribution, and we show the final flavor configuration after the swap network in the two lower panels of Fig.~\ref{fig:pairswaps}.

For the swapped distribution, we find that the average energy of each species is approximately equal and is 
$\bar{E} \approx 14.5$ MeV. 
The number of electron flavor neutrinos and antineutrinos are reduced to $0.38$ and $0.05$, respectively by transformation to other flavors with
equal products of number of neutrinos times antineutrinos in all flavors. The other flavor populations are increased slightly to $0.14$.  While the individual flavor populations have been adjusted, the conserved differences have been respected. This rapid approach to equilibrium can significantly impact the dynamics and chemical evolution of supernovae and neutron star mergers.

\section{Conclusion}
In this work we have provided evidence the generic $\nu-\nu$ coherent forward scattering Hamiltonian is non-integrable, and the resultant level spacing statistics behave equivalently to those of a random matrix 
drawn from a Gaussian orthogonal ensemble. This behavior is generically expected in non-integrable Hamiltonians as there is not an extensive set of conserved quantities which would 
admit degeneracy in the spectrum. Such Hamiltonians display chaotic behavior, and a large category of initial conditions are expected to thermalize in the sense that one-body 
expectation values will obtain some equilibrium value from which they stray only transiently, and which can be predicted from an energy diagonal grand-canonical thermal distribution.

Artificial impositions of symmetry can categorically change the nature of the coherent forward scattering Hamiltonian.  We should only make such simplifying assumptions in the pursuit of 
understanding if doing so does not qualitatively change the behavior of the system we are studying.  This work strongly encourages more careful consideration on how to appropriately take a thermodynamic limit in these systems.

The evidence we present here strongly suggests that late-time one-body expectation values can be obtained \textit{a priori} from a thermal partition function.  
While fully diagonalizing the Hamiltonian for large $N$ is prohibitively expensive, even in the lower dimensional block-decimated invariant subspaces, Monte Carlo methods may be feasibly implemented to evaluate the partition function at larger values of $N$.  Such a scheme may provide a method for feasibly determining the late time one-body flavor expectation values in the fast oscillation regime without explicitly solving the many-body Schr\"{o}dinger equation or numerically diagonalizing the Hamiltonian.

While the time-independent Hamiltonian and initial conditions we consider here are quite simple,
we do not expect the neglected effects, including non-forward scattering, spatially resolved initial states or time dependence in the Hamiltonian, to fundamentally increase the coherence in the system.
In at least some cases a simple classical picture of equilibrium, as determined by assuming an
evolution to an incoherent sum of product states, should provide an accurate approximation to one-body
observables.  In particular, this path integral picture can be used to define an intriguing 
equilibrium distribution for systems of neutrinos and antineutrinos of multiple flavors.

Furthermore in the near future quantum computers may provide an avenue for performing the coherent time evolution of the quantum many-body system (see e.g.~\cite{Hall2021,yeter2022collective,illa2022basic,Illa:2022zgu,Amitrano2023,siwach2023collective} for recent attempts on current generation devices), and may be capable of evaluating expectation values using explicit time evolution for similarly large $N$, which can be compared to statistical partition functions obtained on either classical or quantum computers.  As such, quantum computers may facilitate comparisons between these two predictions. 

Finally, as the number of spins included increases, the block-decimated subspaces of the Hamiltonian grow combinatorially large which we should expect to drive the system even closer to the 
predictions made utilizing the partition function.  
While we have provided evidence which suggests that the equilibration we have observed here will be obtained when $\Hvv$ dominates the evolution, 
a remaining open question is the precise rate at which this thermal equilibrium is achieved.  We generally observe equilibration on a timescale which is on the order of $\sim 10 \mu^{-1} $, however we have not proven that this will occur on such short time scales generically.  The examples we have studied do not display sufficient sensitivity to the system size in the equilibration timescale to make a clear determination of the relationship. We leave a more thorough investigation of the approach to equilibrium to future work.

\acknowledgments{
We thank Vincenzo Cirigliano, Ivan Deutsch, and George Fuller for productive conversations. 
This work was supported by the Quantum Science Center (QSC), a National Quantum Information Science Research Center of the U.S. Department of Energy (DOE) and by the U.S. Department of Energy, Office of Science, Office of Nuclear Physics (NP) contract DE-AC52-06NA25396. H.~D.\ is supported by the US DOE NP grant No.\ DE-SC0017803 at UNM. 
}

\bibliography{refs.bib}

\begin{thebibliography}{83}%
\makeatletter
\providecommand \@ifxundefined [1]{%
 \@ifx{#1\undefined}
}%
\providecommand \@ifnum [1]{%
 \ifnum #1\expandafter \@firstoftwo
 \else \expandafter \@secondoftwo
 \fi
}%
\providecommand \@ifx [1]{%
 \ifx #1\expandafter \@firstoftwo
 \else \expandafter \@secondoftwo
 \fi
}%
\providecommand \natexlab [1]{#1}%
\providecommand \enquote  [1]{``#1''}%
\providecommand \bibnamefont  [1]{#1}%
\providecommand \bibfnamefont [1]{#1}%
\providecommand \citenamefont [1]{#1}%
\providecommand \href@noop [0]{\@secondoftwo}%
\providecommand \href [0]{\begingroup \@sanitize@url \@href}%
\providecommand \@href[1]{\@@startlink{#1}\@@href}%
\providecommand \@@href[1]{\endgroup#1\@@endlink}%
\providecommand \@sanitize@url [0]{\catcode `\\12\catcode `\$12\catcode
  `\&12\catcode `\#12\catcode `\^12\catcode `\_12\catcode `\%12\relax}%
\providecommand \@@startlink[1]{}%
\providecommand \@@endlink[0]{}%
\providecommand \url  [0]{\begingroup\@sanitize@url \@url }%
\providecommand \@url [1]{\endgroup\@href {#1}{\urlprefix }}%
\providecommand \urlprefix  [0]{URL }%
\providecommand \Eprint [0]{\href }%
\providecommand \doibase [0]{http://dx.doi.org/}%
\providecommand \selectlanguage [0]{\@gobble}%
\providecommand \bibinfo  [0]{\@secondoftwo}%
\providecommand \bibfield  [0]{\@secondoftwo}%
\providecommand \translation [1]{[#1]}%
\providecommand \BibitemOpen [0]{}%
\providecommand \bibitemStop [0]{}%
\providecommand \bibitemNoStop [0]{.\EOS\space}%
\providecommand \EOS [0]{\spacefactor3000\relax}%
\providecommand \BibitemShut  [1]{\csname bibitem#1\endcsname}%
\let\auto@bib@innerbib\@empty
\bibitem [{\citenamefont {Bethe}(1990)}]{Bethe_RevModPhys.62.801}%
  \BibitemOpen
  \bibfield  {author} {\bibinfo {author} {\bibfnamefont {H.~A.}\ \bibnamefont
  {Bethe}},\ }\bibfield  {title} {\enquote {\bibinfo {title} {Supernova
  mechanisms},}\ }\href {\doibase 10.1103/RevModPhys.62.801} {\bibfield
  {journal} {\bibinfo  {journal} {Rev. Mod. Phys.}\ }\textbf {\bibinfo {volume}
  {62}},\ \bibinfo {pages} {801--866} (\bibinfo {year} {1990})}\BibitemShut
  {NoStop}%
\bibitem [{\citenamefont {Pantaleone}(1992)}]{pantaleone1992neutrino}%
  \BibitemOpen
  \bibfield  {author} {\bibinfo {author} {\bibfnamefont {James}\ \bibnamefont
  {Pantaleone}},\ }\bibfield  {title} {\enquote {\bibinfo {title} {Neutrino
  oscillations at high densities},}\ }\href@noop {} {\bibfield  {journal}
  {\bibinfo  {journal} {Physics Letters B}\ }\textbf {\bibinfo {volume}
  {287}},\ \bibinfo {pages} {128--132} (\bibinfo {year} {1992})}\BibitemShut
  {NoStop}%
\bibitem [{\citenamefont {Janka}\ \emph {et~al.}(2007)\citenamefont {Janka},
  \citenamefont {Langanke}, \citenamefont {Marek}, \citenamefont
  {Martinez-Pinedo},\ and\ \citenamefont {Mueller}}]{Janka:2006fh}%
  \BibitemOpen
  \bibfield  {author} {\bibinfo {author} {\bibfnamefont {Hans-Thomas}\
  \bibnamefont {Janka}}, \bibinfo {author} {\bibfnamefont {K.}~\bibnamefont
  {Langanke}}, \bibinfo {author} {\bibfnamefont {A.}~\bibnamefont {Marek}},
  \bibinfo {author} {\bibfnamefont {G.}~\bibnamefont {Martinez-Pinedo}}, \ and\
  \bibinfo {author} {\bibfnamefont {B.}~\bibnamefont {Mueller}},\ }\bibfield
  {title} {\enquote {\bibinfo {title} {{Theory of Core-Collapse Supernovae}},}\
  }\href {\doibase 10.1016/j.physrep.2007.02.002} {\bibfield  {journal}
  {\bibinfo  {journal} {Phys. Rept.}\ }\textbf {\bibinfo {volume} {442}},\
  \bibinfo {pages} {38--74} (\bibinfo {year} {2007})},\ \Eprint
  {http://arxiv.org/abs/astro-ph/0612072} {arXiv:astro-ph/0612072 [astro-ph]}
  \BibitemShut {NoStop}%
\bibitem [{\citenamefont {Woosley}\ and\ \citenamefont
  {Janka}(2005)}]{Woosley:2005}%
  \BibitemOpen
  \bibfield  {author} {\bibinfo {author} {\bibfnamefont {Stan}\ \bibnamefont
  {Woosley}}\ and\ \bibinfo {author} {\bibfnamefont {Thomas}\ \bibnamefont
  {Janka}},\ }\bibfield  {title} {\enquote {\bibinfo {title} {The physics of
  core-collapse supernovae},}\ }\href {\doibase 10.1038/nphys172} {\bibfield
  {journal} {\bibinfo  {journal} {Nature Physics}\ }\textbf {\bibinfo {volume}
  {1}},\ \bibinfo {pages} {147--154} (\bibinfo {year} {2005})}\BibitemShut
  {NoStop}%
\bibitem [{\citenamefont {Hoffman}\ \emph {et~al.}(1997)\citenamefont
  {Hoffman}, \citenamefont {Woosley},\ and\ \citenamefont
  {Qian}}]{Hoffman:1997}%
  \BibitemOpen
  \bibfield  {author} {\bibinfo {author} {\bibfnamefont {R.~D.}\ \bibnamefont
  {Hoffman}}, \bibinfo {author} {\bibfnamefont {S.~E.}\ \bibnamefont
  {Woosley}}, \ and\ \bibinfo {author} {\bibfnamefont {Y.‐Z.}\ \bibnamefont
  {Qian}},\ }\bibfield  {title} {\enquote {\bibinfo {title} {Nucleosynthesis in
  neutrino‐driven winds. ii. implications for heavy element synthesis},}\
  }\href {\doibase 10.1086/304181} {\bibfield  {journal} {\bibinfo  {journal}
  {The Astrophysical Journal}\ }\textbf {\bibinfo {volume} {482}},\ \bibinfo
  {pages} {951--962} (\bibinfo {year} {1997})}\BibitemShut {NoStop}%
\bibitem [{\citenamefont {Li}\ and\ \citenamefont {Siegel}(2021)}]{Li:2021vqj}%
  \BibitemOpen
  \bibfield  {author} {\bibinfo {author} {\bibfnamefont {Xinyu}\ \bibnamefont
  {Li}}\ and\ \bibinfo {author} {\bibfnamefont {Daniel~M.}\ \bibnamefont
  {Siegel}},\ }\bibfield  {title} {\enquote {\bibinfo {title} {{Neutrino Fast
  Flavor Conversions in Neutron-Star Postmerger Accretion Disks}},}\ }\href
  {\doibase 10.1103/PhysRevLett.126.251101} {\bibfield  {journal} {\bibinfo
  {journal} {Phys. Rev. Lett.}\ }\textbf {\bibinfo {volume} {126}},\ \bibinfo
  {pages} {251101} (\bibinfo {year} {2021})},\ \Eprint
  {http://arxiv.org/abs/2103.02616} {arXiv:2103.02616 [astro-ph.HE]}
  \BibitemShut {NoStop}%
\bibitem [{\citenamefont {Fern\'andez}\ \emph {et~al.}(2022)\citenamefont
  {Fern\'andez}, \citenamefont {Richers}, \citenamefont {Mulyk},\ and\
  \citenamefont {Fahlman}}]{Fernandez:2022yyv}%
  \BibitemOpen
  \bibfield  {author} {\bibinfo {author} {\bibfnamefont {Rodrigo}\ \bibnamefont
  {Fern\'andez}}, \bibinfo {author} {\bibfnamefont {Sherwood}\ \bibnamefont
  {Richers}}, \bibinfo {author} {\bibfnamefont {Nicole}\ \bibnamefont {Mulyk}},
  \ and\ \bibinfo {author} {\bibfnamefont {Steven}\ \bibnamefont {Fahlman}},\
  }\bibfield  {title} {\enquote {\bibinfo {title} {{Fast flavor instability in
  hypermassive neutron star disk outflows}},}\ }\href {\doibase
  10.1103/PhysRevD.106.103003} {\bibfield  {journal} {\bibinfo  {journal}
  {Phys. Rev. D}\ }\textbf {\bibinfo {volume} {106}},\ \bibinfo {pages}
  {103003} (\bibinfo {year} {2022})},\ \Eprint
  {http://arxiv.org/abs/2207.10680} {arXiv:2207.10680 [astro-ph.HE]}
  \BibitemShut {NoStop}%
\bibitem [{\citenamefont {Sigl}\ and\ \citenamefont
  {Raffelt}(1993)}]{Sigl:1992fn}%
  \BibitemOpen
  \bibfield  {author} {\bibinfo {author} {\bibfnamefont {G.}~\bibnamefont
  {Sigl}}\ and\ \bibinfo {author} {\bibfnamefont {G.}~\bibnamefont {Raffelt}},\
  }\bibfield  {title} {\enquote {\bibinfo {title} {General kinetic description
  of relativistic mixed neutrinos},}\ }\href {\doibase
  https://doi.org/10.1016/0550-3213(93)90175-O} {\bibfield  {journal} {\bibinfo
   {journal} {Nuclear Physics B}\ }\textbf {\bibinfo {volume} {406}},\ \bibinfo
  {pages} {423--451} (\bibinfo {year} {1993})}\BibitemShut {NoStop}%
\bibitem [{\citenamefont {Qian}\ and\ \citenamefont
  {Fuller}(1995{\natexlab{a}})}]{Qian:1994wh}%
  \BibitemOpen
  \bibfield  {author} {\bibinfo {author} {\bibfnamefont {Yong~Zhong}\
  \bibnamefont {Qian}}\ and\ \bibinfo {author} {\bibfnamefont {George~M.}\
  \bibnamefont {Fuller}},\ }\bibfield  {title} {\enquote {\bibinfo {title}
  {{Neutrino-neutrino scattering and matter enhanced neutrino flavor
  transformation in Supernovae}},}\ }\href {\doibase 10.1103/PhysRevD.51.1479}
  {\bibfield  {journal} {\bibinfo  {journal} {Phys. Rev.}\ }\textbf {\bibinfo
  {volume} {D51}},\ \bibinfo {pages} {1479--1494} (\bibinfo {year}
  {1995}{\natexlab{a}})},\ \Eprint {http://arxiv.org/abs/astro-ph/9406073}
  {arXiv:astro-ph/9406073 [astro-ph]} \BibitemShut {NoStop}%
\bibitem [{\citenamefont {Qian}\ and\ \citenamefont
  {Fuller}(1995{\natexlab{b}})}]{Qian:1995}%
  \BibitemOpen
  \bibfield  {author} {\bibinfo {author} {\bibfnamefont {Yong-Zhong}\
  \bibnamefont {Qian}}\ and\ \bibinfo {author} {\bibfnamefont {George~M.}\
  \bibnamefont {Fuller}},\ }\bibfield  {title} {\enquote {\bibinfo {title}
  {Matter-enhanced antineutrino flavor transformation and supernova
  nucleosynthesis},}\ }\href {\doibase 10.1103/physrevd.52.656} {\bibfield
  {journal} {\bibinfo  {journal} {Physical Review D}\ }\textbf {\bibinfo
  {volume} {52}},\ \bibinfo {pages} {656--660} (\bibinfo {year}
  {1995}{\natexlab{b}})}\BibitemShut {NoStop}%
\bibitem [{\citenamefont {Pastor}\ and\ \citenamefont
  {Raffelt}(2002)}]{pastor2002flavor}%
  \BibitemOpen
  \bibfield  {author} {\bibinfo {author} {\bibfnamefont {Sergio}\ \bibnamefont
  {Pastor}}\ and\ \bibinfo {author} {\bibfnamefont {Georg}\ \bibnamefont
  {Raffelt}},\ }\bibfield  {title} {\enquote {\bibinfo {title} {Flavor
  oscillations in the supernova hot bubble region: Nonlinear effects of
  neutrino background},}\ }\href {\doibase 10.1103/PhysRevLett.89.191101}
  {\bibfield  {journal} {\bibinfo  {journal} {Phys. Rev. Lett.}\ }\textbf
  {\bibinfo {volume} {89}},\ \bibinfo {pages} {191101} (\bibinfo {year}
  {2002})}\BibitemShut {NoStop}%
\bibitem [{\citenamefont {Pastor}\ \emph {et~al.}(2002)\citenamefont {Pastor},
  \citenamefont {Raffelt},\ and\ \citenamefont {Semikoz}}]{pastor2002physics}%
  \BibitemOpen
  \bibfield  {author} {\bibinfo {author} {\bibfnamefont {Sergio}\ \bibnamefont
  {Pastor}}, \bibinfo {author} {\bibfnamefont {Georg}\ \bibnamefont {Raffelt}},
  \ and\ \bibinfo {author} {\bibfnamefont {Dmitry~V.}\ \bibnamefont
  {Semikoz}},\ }\bibfield  {title} {\enquote {\bibinfo {title} {Physics of
  synchronized neutrino oscillations caused by self-interactions},}\ }\href
  {\doibase 10.1103/PhysRevD.65.053011} {\bibfield  {journal} {\bibinfo
  {journal} {Phys. Rev. D}\ }\textbf {\bibinfo {volume} {65}},\ \bibinfo
  {pages} {053011} (\bibinfo {year} {2002})}\BibitemShut {NoStop}%
\bibitem [{\citenamefont {Bell}\ \emph {et~al.}(2003)\citenamefont {Bell},
  \citenamefont {Rawlinson},\ and\ \citenamefont {Sawyer}}]{Bell:2003mg}%
  \BibitemOpen
  \bibfield  {author} {\bibinfo {author} {\bibfnamefont {Nicole~F.}\
  \bibnamefont {Bell}}, \bibinfo {author} {\bibfnamefont {Andrew~A.}\
  \bibnamefont {Rawlinson}}, \ and\ \bibinfo {author} {\bibfnamefont {R.~F.}\
  \bibnamefont {Sawyer}},\ }\bibfield  {title} {\enquote {\bibinfo {title}
  {{Speedup through entanglement: Many body effects in neutrino processes}},}\
  }\href {\doibase 10.1016/j.physletb.2003.08.035} {\bibfield  {journal}
  {\bibinfo  {journal} {Phys. Lett. B}\ }\textbf {\bibinfo {volume} {573}},\
  \bibinfo {pages} {86--93} (\bibinfo {year} {2003})},\ \Eprint
  {http://arxiv.org/abs/hep-ph/0304082} {arXiv:hep-ph/0304082} \BibitemShut
  {NoStop}%
\bibitem [{\citenamefont {Sawyer}(2004)}]{sawyer2004classical}%
  \BibitemOpen
  \bibfield  {author} {\bibinfo {author} {\bibfnamefont {R.F.}\ \bibnamefont
  {Sawyer}},\ }\bibfield  {title} {\enquote {\bibinfo {title} {{'Classical'
  instabilities and 'quantum' speed-up in the evolution of neutrino clouds}},}\
  }\href@noop {} {\  (\bibinfo {year} {2004})},\ \Eprint
  {http://arxiv.org/abs/hep-ph/0408265} {arXiv:hep-ph/0408265} \BibitemShut
  {NoStop}%
\bibitem [{\citenamefont {Balantekin}\ and\ \citenamefont
  {Pehlivan}(2007)}]{Balantekin:2006tg}%
  \BibitemOpen
  \bibfield  {author} {\bibinfo {author} {\bibfnamefont {A.B.}\ \bibnamefont
  {Balantekin}}\ and\ \bibinfo {author} {\bibfnamefont {Y.}~\bibnamefont
  {Pehlivan}},\ }\bibfield  {title} {\enquote {\bibinfo {title}
  {{Neutrino-Neutrino Interactions and Flavor Mixing in Dense Matter}},}\
  }\href {\doibase 10.1088/0954-3899/34/1/004} {\bibfield  {journal} {\bibinfo
  {journal} {J. Phys. G}\ }\textbf {\bibinfo {volume} {34}},\ \bibinfo {pages}
  {47--66} (\bibinfo {year} {2007})},\ \Eprint
  {http://arxiv.org/abs/astro-ph/0607527} {arXiv:astro-ph/0607527} \BibitemShut
  {NoStop}%
\bibitem [{\citenamefont {Duan}\ \emph
  {et~al.}(2006{\natexlab{a}})\citenamefont {Duan}, \citenamefont {Fuller},
  \citenamefont {Carlson},\ and\ \citenamefont {Qian}}]{Duan:2006jv}%
  \BibitemOpen
  \bibfield  {author} {\bibinfo {author} {\bibfnamefont {Huaiyu}\ \bibnamefont
  {Duan}}, \bibinfo {author} {\bibfnamefont {George~M.}\ \bibnamefont
  {Fuller}}, \bibinfo {author} {\bibfnamefont {J.}~\bibnamefont {Carlson}}, \
  and\ \bibinfo {author} {\bibfnamefont {Yong-Zhong}\ \bibnamefont {Qian}},\
  }\bibfield  {title} {\enquote {\bibinfo {title} {{Coherent Development of
  Neutrino Flavor in the Supernova Environment}},}\ }\href {\doibase
  10.1103/PhysRevLett.97.241101} {\bibfield  {journal} {\bibinfo  {journal}
  {Phys. Rev. Lett.}\ }\textbf {\bibinfo {volume} {97}},\ \bibinfo {pages}
  {241101} (\bibinfo {year} {2006}{\natexlab{a}})},\ \Eprint
  {http://arxiv.org/abs/astro-ph/0608050} {arXiv:astro-ph/0608050} \BibitemShut
  {NoStop}%
\bibitem [{\citenamefont {Raffelt}\ and\ \citenamefont
  {Smirnov}(2007{\natexlab{a}})}]{Raffelt:2007cb}%
  \BibitemOpen
  \bibfield  {author} {\bibinfo {author} {\bibfnamefont {Georg~G.}\
  \bibnamefont {Raffelt}}\ and\ \bibinfo {author} {\bibfnamefont {Alexei~Yu.}\
  \bibnamefont {Smirnov}},\ }\bibfield  {title} {\enquote {\bibinfo {title}
  {{Self-induced spectral splits in supernova neutrino fluxes}},}\ }\href
  {\doibase 10.1103/PhysRevD.76.081301} {\bibfield  {journal} {\bibinfo
  {journal} {Phys. Rev. D}\ }\textbf {\bibinfo {volume} {76}},\ \bibinfo
  {pages} {081301} (\bibinfo {year} {2007}{\natexlab{a}})},\ \bibinfo {note}
  {[Erratum: Phys.Rev.D 77, 029903 (2008)]},\ \Eprint
  {http://arxiv.org/abs/0705.1830} {arXiv:0705.1830 [hep-ph]} \BibitemShut
  {NoStop}%
\bibitem [{\citenamefont {Raffelt}\ and\ \citenamefont
  {Smirnov}(2007{\natexlab{b}})}]{Raffelt:2007xt}%
  \BibitemOpen
  \bibfield  {author} {\bibinfo {author} {\bibfnamefont {Georg~G.}\
  \bibnamefont {Raffelt}}\ and\ \bibinfo {author} {\bibfnamefont {Alexei~Yu.}\
  \bibnamefont {Smirnov}},\ }\bibfield  {title} {\enquote {\bibinfo {title}
  {{Adiabaticity and spectral splits in collective neutrino
  transformations}},}\ }\href {\doibase 10.1103/PhysRevD.76.125008} {\bibfield
  {journal} {\bibinfo  {journal} {Phys. Rev. D}\ }\textbf {\bibinfo {volume}
  {76}},\ \bibinfo {pages} {125008} (\bibinfo {year} {2007}{\natexlab{b}})},\
  \Eprint {http://arxiv.org/abs/0709.4641} {arXiv:0709.4641 [hep-ph]}
  \BibitemShut {NoStop}%
\bibitem [{\citenamefont {Sawyer}(2005)}]{PhysRevD.72.045003}%
  \BibitemOpen
  \bibfield  {author} {\bibinfo {author} {\bibfnamefont {R.~F.}\ \bibnamefont
  {Sawyer}},\ }\bibfield  {title} {\enquote {\bibinfo {title} {Speed-up of
  neutrino transformations in a supernova environment},}\ }\href {\doibase
  10.1103/PhysRevD.72.045003} {\bibfield  {journal} {\bibinfo  {journal} {Phys.
  Rev. D}\ }\textbf {\bibinfo {volume} {72}},\ \bibinfo {pages} {045003}
  (\bibinfo {year} {2005})}\BibitemShut {NoStop}%
\bibitem [{\citenamefont {Sawyer}(2009)}]{PhysRevD.79.105003}%
  \BibitemOpen
  \bibfield  {author} {\bibinfo {author} {\bibfnamefont {R.~F.}\ \bibnamefont
  {Sawyer}},\ }\bibfield  {title} {\enquote {\bibinfo {title} {Multiangle
  instability in dense neutrino systems},}\ }\href {\doibase
  10.1103/PhysRevD.79.105003} {\bibfield  {journal} {\bibinfo  {journal} {Phys.
  Rev. D}\ }\textbf {\bibinfo {volume} {79}},\ \bibinfo {pages} {105003}
  (\bibinfo {year} {2009})}\BibitemShut {NoStop}%
\bibitem [{\citenamefont {Sawyer}(2016)}]{PhysRevLett.116.081101}%
  \BibitemOpen
  \bibfield  {author} {\bibinfo {author} {\bibfnamefont {R.~F.}\ \bibnamefont
  {Sawyer}},\ }\bibfield  {title} {\enquote {\bibinfo {title} {Neutrino cloud
  instabilities just above the neutrino sphere of a supernova},}\ }\href
  {\doibase 10.1103/PhysRevLett.116.081101} {\bibfield  {journal} {\bibinfo
  {journal} {Phys. Rev. Lett.}\ }\textbf {\bibinfo {volume} {116}},\ \bibinfo
  {pages} {081101} (\bibinfo {year} {2016})}\BibitemShut {NoStop}%
\bibitem [{\citenamefont {Chakraborty}\ \emph {et~al.}(2016)\citenamefont
  {Chakraborty}, \citenamefont {Hansen}, \citenamefont {Izaguirre},\ and\
  \citenamefont {Raffelt}}]{chakraborty2016self}%
  \BibitemOpen
  \bibfield  {author} {\bibinfo {author} {\bibfnamefont {Sovan}\ \bibnamefont
  {Chakraborty}}, \bibinfo {author} {\bibfnamefont {Rasmus~Sloth}\ \bibnamefont
  {Hansen}}, \bibinfo {author} {\bibfnamefont {Ignacio}\ \bibnamefont
  {Izaguirre}}, \ and\ \bibinfo {author} {\bibfnamefont {Georg~G}\ \bibnamefont
  {Raffelt}},\ }\bibfield  {title} {\enquote {\bibinfo {title} {Self-induced
  neutrino flavor conversion without flavor mixing},}\ }\href@noop {}
  {\bibfield  {journal} {\bibinfo  {journal} {Journal of Cosmology and
  Astroparticle Physics}\ }\textbf {\bibinfo {volume} {2016}},\ \bibinfo
  {pages} {042} (\bibinfo {year} {2016})}\BibitemShut {NoStop}%
\bibitem [{\citenamefont {Izaguirre}\ \emph {et~al.}(2017)\citenamefont
  {Izaguirre}, \citenamefont {Raffelt},\ and\ \citenamefont
  {Tamborra}}]{PhysRevLett.118.021101}%
  \BibitemOpen
  \bibfield  {author} {\bibinfo {author} {\bibfnamefont {Ignacio}\ \bibnamefont
  {Izaguirre}}, \bibinfo {author} {\bibfnamefont {Georg}\ \bibnamefont
  {Raffelt}}, \ and\ \bibinfo {author} {\bibfnamefont {Irene}\ \bibnamefont
  {Tamborra}},\ }\bibfield  {title} {\enquote {\bibinfo {title} {Fast pairwise
  conversion of supernova neutrinos: A dispersion relation approach},}\ }\href
  {\doibase 10.1103/PhysRevLett.118.021101} {\bibfield  {journal} {\bibinfo
  {journal} {Phys. Rev. Lett.}\ }\textbf {\bibinfo {volume} {118}},\ \bibinfo
  {pages} {021101} (\bibinfo {year} {2017})}\BibitemShut {NoStop}%
\bibitem [{\citenamefont {Dasgupta}\ \emph {et~al.}(2017)\citenamefont
  {Dasgupta}, \citenamefont {Mirizzi},\ and\ \citenamefont
  {Sen}}]{dasgupta2017fast}%
  \BibitemOpen
  \bibfield  {author} {\bibinfo {author} {\bibfnamefont {Basudeb}\ \bibnamefont
  {Dasgupta}}, \bibinfo {author} {\bibfnamefont {Alessandro}\ \bibnamefont
  {Mirizzi}}, \ and\ \bibinfo {author} {\bibfnamefont {Manibrata}\ \bibnamefont
  {Sen}},\ }\bibfield  {title} {\enquote {\bibinfo {title} {Fast neutrino
  flavor conversions near the supernova core with realistic flavor-dependent
  angular distributions},}\ }\href@noop {} {\bibfield  {journal} {\bibinfo
  {journal} {Journal of Cosmology and Astroparticle Physics}\ }\textbf
  {\bibinfo {volume} {2017}},\ \bibinfo {pages} {019} (\bibinfo {year}
  {2017})}\BibitemShut {NoStop}%
\bibitem [{\citenamefont {Wu}\ and\ \citenamefont
  {Tamborra}(2017)}]{PhysRevD.95.103007}%
  \BibitemOpen
  \bibfield  {author} {\bibinfo {author} {\bibfnamefont {Meng-Ru}\ \bibnamefont
  {Wu}}\ and\ \bibinfo {author} {\bibfnamefont {Irene}\ \bibnamefont
  {Tamborra}},\ }\bibfield  {title} {\enquote {\bibinfo {title} {Fast neutrino
  conversions: Ubiquitous in compact binary merger remnants},}\ }\href
  {\doibase 10.1103/PhysRevD.95.103007} {\bibfield  {journal} {\bibinfo
  {journal} {Phys. Rev. D}\ }\textbf {\bibinfo {volume} {95}},\ \bibinfo
  {pages} {103007} (\bibinfo {year} {2017})}\BibitemShut {NoStop}%
\bibitem [{\citenamefont {Abbar}\ and\ \citenamefont
  {Duan}(2018)}]{PhysRevD.98.043014}%
  \BibitemOpen
  \bibfield  {author} {\bibinfo {author} {\bibfnamefont {Sajad}\ \bibnamefont
  {Abbar}}\ and\ \bibinfo {author} {\bibfnamefont {Huaiyu}\ \bibnamefont
  {Duan}},\ }\bibfield  {title} {\enquote {\bibinfo {title} {Fast neutrino
  flavor conversion: Roles of dense matter and spectrum crossing},}\ }\href
  {\doibase 10.1103/PhysRevD.98.043014} {\bibfield  {journal} {\bibinfo
  {journal} {Phys. Rev. D}\ }\textbf {\bibinfo {volume} {98}},\ \bibinfo
  {pages} {043014} (\bibinfo {year} {2018})}\BibitemShut {NoStop}%
\bibitem [{\citenamefont {Dasgupta}\ \emph {et~al.}(2018)\citenamefont
  {Dasgupta}, \citenamefont {Mirizzi},\ and\ \citenamefont
  {Sen}}]{PhysRevD.98.103001}%
  \BibitemOpen
  \bibfield  {author} {\bibinfo {author} {\bibfnamefont {Basudeb}\ \bibnamefont
  {Dasgupta}}, \bibinfo {author} {\bibfnamefont {Alessandro}\ \bibnamefont
  {Mirizzi}}, \ and\ \bibinfo {author} {\bibfnamefont {Manibrata}\ \bibnamefont
  {Sen}},\ }\bibfield  {title} {\enquote {\bibinfo {title} {Simple method of
  diagnosing fast flavor conversions of supernova neutrinos},}\ }\href
  {\doibase 10.1103/PhysRevD.98.103001} {\bibfield  {journal} {\bibinfo
  {journal} {Phys. Rev. D}\ }\textbf {\bibinfo {volume} {98}},\ \bibinfo
  {pages} {103001} (\bibinfo {year} {2018})}\BibitemShut {NoStop}%
\bibitem [{\citenamefont {Airen}\ \emph {et~al.}(2018)\citenamefont {Airen},
  \citenamefont {Capozzi}, \citenamefont {Chakraborty}, \citenamefont
  {Dasgupta}, \citenamefont {Raffelt},\ and\ \citenamefont
  {Stirner}}]{airen2018normal}%
  \BibitemOpen
  \bibfield  {author} {\bibinfo {author} {\bibfnamefont {Sagar}\ \bibnamefont
  {Airen}}, \bibinfo {author} {\bibfnamefont {Francesco}\ \bibnamefont
  {Capozzi}}, \bibinfo {author} {\bibfnamefont {Sovan}\ \bibnamefont
  {Chakraborty}}, \bibinfo {author} {\bibfnamefont {Basudeb}\ \bibnamefont
  {Dasgupta}}, \bibinfo {author} {\bibfnamefont {Georg}\ \bibnamefont
  {Raffelt}}, \ and\ \bibinfo {author} {\bibfnamefont {Tobias}\ \bibnamefont
  {Stirner}},\ }\bibfield  {title} {\enquote {\bibinfo {title} {Normal-mode
  analysis for collective neutrino oscillations},}\ }\href@noop {} {\bibfield
  {journal} {\bibinfo  {journal} {Journal of Cosmology and Astroparticle
  Physics}\ }\textbf {\bibinfo {volume} {2018}},\ \bibinfo {pages} {019}
  (\bibinfo {year} {2018})}\BibitemShut {NoStop}%
\bibitem [{\citenamefont {Martin}\ \emph {et~al.}(2020)\citenamefont {Martin},
  \citenamefont {Yi},\ and\ \citenamefont {Duan}}]{Martin:2019gxb}%
  \BibitemOpen
  \bibfield  {author} {\bibinfo {author} {\bibfnamefont {Joshua~D.}\
  \bibnamefont {Martin}}, \bibinfo {author} {\bibfnamefont {Changhao}\
  \bibnamefont {Yi}}, \ and\ \bibinfo {author} {\bibfnamefont {Huaiyu}\
  \bibnamefont {Duan}},\ }\bibfield  {title} {\enquote {\bibinfo {title}
  {{Dynamic fast flavor oscillation waves in dense neutrino gases}},}\ }\href
  {\doibase 10.1016/j.physletb.2019.135088} {\bibfield  {journal} {\bibinfo
  {journal} {Phys. Lett.}\ }\textbf {\bibinfo {volume} {B800}},\ \bibinfo
  {pages} {135088} (\bibinfo {year} {2020})},\ \Eprint
  {http://arxiv.org/abs/1909.05225} {arXiv:1909.05225 [hep-ph]} \BibitemShut
  {NoStop}%
\bibitem [{\citenamefont {Johns}\ \emph {et~al.}(2020)\citenamefont {Johns},
  \citenamefont {Nagakura}, \citenamefont {Fuller},\ and\ \citenamefont
  {Burrows}}]{PhysRevD.101.043009}%
  \BibitemOpen
  \bibfield  {author} {\bibinfo {author} {\bibfnamefont {Lucas}\ \bibnamefont
  {Johns}}, \bibinfo {author} {\bibfnamefont {Hiroki}\ \bibnamefont
  {Nagakura}}, \bibinfo {author} {\bibfnamefont {George~M.}\ \bibnamefont
  {Fuller}}, \ and\ \bibinfo {author} {\bibfnamefont {Adam}\ \bibnamefont
  {Burrows}},\ }\bibfield  {title} {\enquote {\bibinfo {title} {Neutrino
  oscillations in supernovae: Angular moments and fast instabilities},}\ }\href
  {\doibase 10.1103/PhysRevD.101.043009} {\bibfield  {journal} {\bibinfo
  {journal} {Phys. Rev. D}\ }\textbf {\bibinfo {volume} {101}},\ \bibinfo
  {pages} {043009} (\bibinfo {year} {2020})}\BibitemShut {NoStop}%
\bibitem [{\citenamefont {Roggero}\ \emph {et~al.}(2022)\citenamefont
  {Roggero}, \citenamefont {Rrapaj},\ and\ \citenamefont
  {Xiong}}]{Roggero2022}%
  \BibitemOpen
  \bibfield  {author} {\bibinfo {author} {\bibfnamefont {Alessandro}\
  \bibnamefont {Roggero}}, \bibinfo {author} {\bibfnamefont {Ermal}\
  \bibnamefont {Rrapaj}}, \ and\ \bibinfo {author} {\bibfnamefont {Zewei}\
  \bibnamefont {Xiong}},\ }\bibfield  {title} {\enquote {\bibinfo {title}
  {Entanglement and correlations in fast collective neutrino flavor
  oscillations},}\ }\href {\doibase 10.1103/PhysRevD.106.043022} {\bibfield
  {journal} {\bibinfo  {journal} {Phys. Rev. D}\ }\textbf {\bibinfo {volume}
  {106}},\ \bibinfo {pages} {043022} (\bibinfo {year} {2022})}\BibitemShut
  {NoStop}%
\bibitem [{\citenamefont {Bhattacharyya}\ and\ \citenamefont
  {Dasgupta}(2022)}]{PhysRevD.106.103039}%
  \BibitemOpen
  \bibfield  {author} {\bibinfo {author} {\bibfnamefont {Soumya}\ \bibnamefont
  {Bhattacharyya}}\ and\ \bibinfo {author} {\bibfnamefont {Basudeb}\
  \bibnamefont {Dasgupta}},\ }\bibfield  {title} {\enquote {\bibinfo {title}
  {Elaborating the ultimate fate of fast collective neutrino flavor
  oscillations},}\ }\href {\doibase 10.1103/PhysRevD.106.103039} {\bibfield
  {journal} {\bibinfo  {journal} {Phys. Rev. D}\ }\textbf {\bibinfo {volume}
  {106}},\ \bibinfo {pages} {103039} (\bibinfo {year} {2022})}\BibitemShut
  {NoStop}%
\bibitem [{\citenamefont {Xiong}\ and\ \citenamefont
  {Qian}(2021)}]{Xiong:2021dex}%
  \BibitemOpen
  \bibfield  {author} {\bibinfo {author} {\bibfnamefont {Zewei}\ \bibnamefont
  {Xiong}}\ and\ \bibinfo {author} {\bibfnamefont {Yong-Zhong}\ \bibnamefont
  {Qian}},\ }\bibfield  {title} {\enquote {\bibinfo {title} {{Stationary
  solutions for fast flavor oscillations of a homogeneous dense neutrino
  gas}},}\ }\href {\doibase 10.1016/j.physletb.2021.136550} {\bibfield
  {journal} {\bibinfo  {journal} {Phys. Lett. B}\ }\textbf {\bibinfo {volume}
  {820}},\ \bibinfo {pages} {136550} (\bibinfo {year} {2021})},\ \Eprint
  {http://arxiv.org/abs/2104.05618} {arXiv:2104.05618 [astro-ph.HE]}
  \BibitemShut {NoStop}%
\bibitem [{\citenamefont {Volpe}(2023)}]{Volpe:2023met}%
  \BibitemOpen
  \bibfield  {author} {\bibinfo {author} {\bibfnamefont {Maria~Cristina}\
  \bibnamefont {Volpe}},\ }\bibfield  {title} {\enquote {\bibinfo {title}
  {{Neutrinos from dense: flavor mechanisms, theoretical approaches,
  observations, new directions}},}\ }\href@noop {} {\  (\bibinfo {year}
  {2023})},\ \Eprint {http://arxiv.org/abs/2301.11814} {arXiv:2301.11814
  [hep-ph]} \BibitemShut {NoStop}%
\bibitem [{\citenamefont {Wolfenstein}(1978)}]{wolfenstein_PhysRevD.17.2369}%
  \BibitemOpen
  \bibfield  {author} {\bibinfo {author} {\bibfnamefont {L.}~\bibnamefont
  {Wolfenstein}},\ }\bibfield  {title} {\enquote {\bibinfo {title} {Neutrino
  oscillations in matter},}\ }\href {\doibase 10.1103/PhysRevD.17.2369}
  {\bibfield  {journal} {\bibinfo  {journal} {Phys. Rev. D}\ }\textbf {\bibinfo
  {volume} {17}},\ \bibinfo {pages} {2369--2374} (\bibinfo {year}
  {1978})}\BibitemShut {NoStop}%
\bibitem [{\citenamefont {Mikheev}\ and\ \citenamefont
  {Smirnov}(1985)}]{mikheev1985resonance}%
  \BibitemOpen
  \bibfield  {author} {\bibinfo {author} {\bibfnamefont {SP}~\bibnamefont
  {Mikheev}}\ and\ \bibinfo {author} {\bibfnamefont {A~Yu}\ \bibnamefont
  {Smirnov}},\ }\bibfield  {title} {\enquote {\bibinfo {title} {Resonance
  amplification of oscillations in matter and spectroscopy of solar
  neutrinos},}\ }\href@noop {} {\bibfield  {journal} {\bibinfo  {journal}
  {Yadernaya Fizika}\ }\textbf {\bibinfo {volume} {42}},\ \bibinfo {pages}
  {1441--1448} (\bibinfo {year} {1985})}\BibitemShut {NoStop}%
\bibitem [{\citenamefont {Friedland}\ \emph {et~al.}(2006)\citenamefont
  {Friedland}, \citenamefont {McKellar},\ and\ \citenamefont
  {Okuniewicz}}]{Friedland2006}%
  \BibitemOpen
  \bibfield  {author} {\bibinfo {author} {\bibfnamefont {Alexander}\
  \bibnamefont {Friedland}}, \bibinfo {author} {\bibfnamefont {Bruce H.~J.}\
  \bibnamefont {McKellar}}, \ and\ \bibinfo {author} {\bibfnamefont {Ivona}\
  \bibnamefont {Okuniewicz}},\ }\bibfield  {title} {\enquote {\bibinfo {title}
  {Construction and analysis of a simplified many-body neutrino model},}\
  }\href {\doibase 10.1103/PhysRevD.73.093002} {\bibfield  {journal} {\bibinfo
  {journal} {Phys. Rev. D}\ }\textbf {\bibinfo {volume} {73}},\ \bibinfo
  {pages} {093002} (\bibinfo {year} {2006})}\BibitemShut {NoStop}%
\bibitem [{\citenamefont {McKellar}\ \emph {et~al.}(2009)\citenamefont
  {McKellar}, \citenamefont {Okuniewicz},\ and\ \citenamefont
  {Quach}}]{PhysRevD.80.013011}%
  \BibitemOpen
  \bibfield  {author} {\bibinfo {author} {\bibfnamefont {Bruce H.~J.}\
  \bibnamefont {McKellar}}, \bibinfo {author} {\bibfnamefont {Ivona}\
  \bibnamefont {Okuniewicz}}, \ and\ \bibinfo {author} {\bibfnamefont {James}\
  \bibnamefont {Quach}},\ }\bibfield  {title} {\enquote {\bibinfo {title}
  {Non-boltzmann behavior in models of interacting neutrinos},}\ }\href
  {\doibase 10.1103/PhysRevD.80.013011} {\bibfield  {journal} {\bibinfo
  {journal} {Phys. Rev. D}\ }\textbf {\bibinfo {volume} {80}},\ \bibinfo
  {pages} {013011} (\bibinfo {year} {2009})}\BibitemShut {NoStop}%
\bibitem [{\citenamefont {Pehlivan}\ \emph {et~al.}(2011)\citenamefont
  {Pehlivan}, \citenamefont {Balantekin}, \citenamefont {Kajino},\ and\
  \citenamefont {Yoshida}}]{Pehlivan2011}%
  \BibitemOpen
  \bibfield  {author} {\bibinfo {author} {\bibfnamefont {Y.}~\bibnamefont
  {Pehlivan}}, \bibinfo {author} {\bibfnamefont {A.~B.}\ \bibnamefont
  {Balantekin}}, \bibinfo {author} {\bibfnamefont {Toshitaka}\ \bibnamefont
  {Kajino}}, \ and\ \bibinfo {author} {\bibfnamefont {Takashi}\ \bibnamefont
  {Yoshida}},\ }\bibfield  {title} {\enquote {\bibinfo {title} {Invariants of
  collective neutrino oscillations},}\ }\href {\doibase
  10.1103/PhysRevD.84.065008} {\bibfield  {journal} {\bibinfo  {journal} {Phys.
  Rev. D}\ }\textbf {\bibinfo {volume} {84}},\ \bibinfo {pages} {065008}
  (\bibinfo {year} {2011})}\BibitemShut {NoStop}%
\bibitem [{\citenamefont {Birol}\ \emph {et~al.}(2018)\citenamefont {Birol},
  \citenamefont {Pehlivan}, \citenamefont {Balantekin},\ and\ \citenamefont
  {Kajino}}]{Birol:2018qhx}%
  \BibitemOpen
  \bibfield  {author} {\bibinfo {author} {\bibfnamefont {Savas}\ \bibnamefont
  {Birol}}, \bibinfo {author} {\bibfnamefont {Y.}~\bibnamefont {Pehlivan}},
  \bibinfo {author} {\bibfnamefont {A.B.}\ \bibnamefont {Balantekin}}, \ and\
  \bibinfo {author} {\bibfnamefont {T.}~\bibnamefont {Kajino}},\ }\bibfield
  {title} {\enquote {\bibinfo {title} {{Neutrino Spectral Split in the Exact
  Many Body Formalism}},}\ }\href {\doibase 10.1103/PhysRevD.98.083002}
  {\bibfield  {journal} {\bibinfo  {journal} {Phys. Rev. D}\ }\textbf {\bibinfo
  {volume} {98}},\ \bibinfo {pages} {083002} (\bibinfo {year} {2018})},\
  \Eprint {http://arxiv.org/abs/1805.11767} {arXiv:1805.11767 [astro-ph.HE]}
  \BibitemShut {NoStop}%
\bibitem [{\citenamefont {Patwardhan}\ \emph {et~al.}(2019)\citenamefont
  {Patwardhan}, \citenamefont {Cervia},\ and\ \citenamefont
  {Balantekin}}]{Patwardhan2019}%
  \BibitemOpen
  \bibfield  {author} {\bibinfo {author} {\bibfnamefont {Amol~V.}\ \bibnamefont
  {Patwardhan}}, \bibinfo {author} {\bibfnamefont {Michael~J.}\ \bibnamefont
  {Cervia}}, \ and\ \bibinfo {author} {\bibfnamefont {A.~Baha}\ \bibnamefont
  {Balantekin}},\ }\bibfield  {title} {\enquote {\bibinfo {title} {Eigenvalues
  and eigenstates of the many-body collective neutrino oscillation problem},}\
  }\href {\doibase 10.1103/PhysRevD.99.123013} {\bibfield  {journal} {\bibinfo
  {journal} {Phys. Rev. D}\ }\textbf {\bibinfo {volume} {99}},\ \bibinfo
  {pages} {123013} (\bibinfo {year} {2019})}\BibitemShut {NoStop}%
\bibitem [{\citenamefont {Patwardhan}\ \emph {et~al.}(2021)\citenamefont
  {Patwardhan}, \citenamefont {Cervia},\ and\ \citenamefont
  {Balantekin}}]{patwardhan2021spectral}%
  \BibitemOpen
  \bibfield  {author} {\bibinfo {author} {\bibfnamefont {Amol~V.}\ \bibnamefont
  {Patwardhan}}, \bibinfo {author} {\bibfnamefont {Michael~J.}\ \bibnamefont
  {Cervia}}, \ and\ \bibinfo {author} {\bibfnamefont {A.~B.}\ \bibnamefont
  {Balantekin}},\ }\bibfield  {title} {\enquote {\bibinfo {title} {Spectral
  splits and entanglement entropy in collective neutrino oscillations},}\
  }\href {\doibase 10.1103/PhysRevD.104.123035} {\bibfield  {journal} {\bibinfo
   {journal} {Phys. Rev. D}\ }\textbf {\bibinfo {volume} {104}},\ \bibinfo
  {pages} {123035} (\bibinfo {year} {2021})}\BibitemShut {NoStop}%
\bibitem [{\citenamefont {Martin}\ \emph {et~al.}(2022)\citenamefont {Martin},
  \citenamefont {Roggero}, \citenamefont {Duan}, \citenamefont {Carlson},\ and\
  \citenamefont {Cirigliano}}]{Martin:2021bri}%
  \BibitemOpen
  \bibfield  {author} {\bibinfo {author} {\bibfnamefont {Joshua~D.}\
  \bibnamefont {Martin}}, \bibinfo {author} {\bibfnamefont {A.}~\bibnamefont
  {Roggero}}, \bibinfo {author} {\bibfnamefont {Huaiyu}\ \bibnamefont {Duan}},
  \bibinfo {author} {\bibfnamefont {J.}~\bibnamefont {Carlson}}, \ and\
  \bibinfo {author} {\bibfnamefont {V.}~\bibnamefont {Cirigliano}},\ }\bibfield
   {title} {\enquote {\bibinfo {title} {{Classical and quantum evolution in a
  simple coherent neutrino problem}},}\ }\href {\doibase
  10.1103/PhysRevD.105.083020} {\bibfield  {journal} {\bibinfo  {journal}
  {Phys. Rev. D}\ }\textbf {\bibinfo {volume} {105}},\ \bibinfo {pages}
  {083020} (\bibinfo {year} {2022})},\ \Eprint
  {http://arxiv.org/abs/2112.12686} {arXiv:2112.12686 [hep-ph]} \BibitemShut
  {NoStop}%
\bibitem [{\citenamefont {Illa}\ and\ \citenamefont
  {Savage}(2023)}]{Illa:2022zgu}%
  \BibitemOpen
  \bibfield  {author} {\bibinfo {author} {\bibfnamefont {Marc}\ \bibnamefont
  {Illa}}\ and\ \bibinfo {author} {\bibfnamefont {Martin~J.}\ \bibnamefont
  {Savage}},\ }\bibfield  {title} {\enquote {\bibinfo {title} {Multi-neutrino
  entanglement and correlations in dense neutrino systems},}\ }\href {\doibase
  10.1103/PhysRevLett.130.221003} {\bibfield  {journal} {\bibinfo  {journal}
  {Phys. Rev. Lett.}\ }\textbf {\bibinfo {volume} {130}},\ \bibinfo {pages}
  {221003} (\bibinfo {year} {2023})}\BibitemShut {NoStop}%
\bibitem [{\citenamefont {Xiong}(2022)}]{Xiong:2021evk}%
  \BibitemOpen
  \bibfield  {author} {\bibinfo {author} {\bibfnamefont {Zewei}\ \bibnamefont
  {Xiong}},\ }\bibfield  {title} {\enquote {\bibinfo {title} {{Many-body
  effects of collective neutrino oscillations}},}\ }\href {\doibase
  10.1103/PhysRevD.105.103002} {\bibfield  {journal} {\bibinfo  {journal}
  {Phys. Rev. D}\ }\textbf {\bibinfo {volume} {105}},\ \bibinfo {pages}
  {103002} (\bibinfo {year} {2022})},\ \Eprint
  {http://arxiv.org/abs/2111.00437} {arXiv:2111.00437 [astro-ph.HE]}
  \BibitemShut {NoStop}%
\bibitem [{\citenamefont {Shalgar}\ and\ \citenamefont
  {Tamborra}(2023)}]{shalgar2023evidence}%
  \BibitemOpen
  \bibfield  {author} {\bibinfo {author} {\bibfnamefont {Shashank}\
  \bibnamefont {Shalgar}}\ and\ \bibinfo {author} {\bibfnamefont {Irene}\
  \bibnamefont {Tamborra}},\ }\bibfield  {title} {\enquote {\bibinfo {title}
  {Do we have enough evidence to invalidate the mean-field approximation
  adopted to model collective neutrino oscillations?}}\ }\href {\doibase
  10.1103/PhysRevD.107.123004} {\bibfield  {journal} {\bibinfo  {journal}
  {Phys. Rev. D}\ }\textbf {\bibinfo {volume} {107}},\ \bibinfo {pages}
  {123004} (\bibinfo {year} {2023})}\BibitemShut {NoStop}%
\bibitem [{\citenamefont {Rrapaj}(2020)}]{Rrapaj2020}%
  \BibitemOpen
  \bibfield  {author} {\bibinfo {author} {\bibfnamefont {Ermal}\ \bibnamefont
  {Rrapaj}},\ }\bibfield  {title} {\enquote {\bibinfo {title} {Exact solution
  of multiangle quantum many-body collective neutrino-flavor oscillations},}\
  }\href {\doibase 10.1103/PhysRevC.101.065805} {\bibfield  {journal} {\bibinfo
   {journal} {Phys. Rev. C}\ }\textbf {\bibinfo {volume} {101}},\ \bibinfo
  {pages} {065805} (\bibinfo {year} {2020})}\BibitemShut {NoStop}%
\bibitem [{\citenamefont {Roggero}(2021{\natexlab{a}})}]{Roggero2021a}%
  \BibitemOpen
  \bibfield  {author} {\bibinfo {author} {\bibfnamefont {Alessandro}\
  \bibnamefont {Roggero}},\ }\bibfield  {title} {\enquote {\bibinfo {title}
  {Entanglement and many-body effects in collective neutrino oscillations},}\
  }\href {\doibase 10.1103/PhysRevD.104.103016} {\bibfield  {journal} {\bibinfo
   {journal} {Phys. Rev. D}\ }\textbf {\bibinfo {volume} {104}},\ \bibinfo
  {pages} {103016} (\bibinfo {year} {2021}{\natexlab{a}})}\BibitemShut
  {NoStop}%
\bibitem [{\citenamefont {Roggero}(2021{\natexlab{b}})}]{Roggero2021b}%
  \BibitemOpen
  \bibfield  {author} {\bibinfo {author} {\bibfnamefont {Alessandro}\
  \bibnamefont {Roggero}},\ }\bibfield  {title} {\enquote {\bibinfo {title}
  {Dynamical phase transitions in models of collective neutrino
  oscillations},}\ }\href {\doibase 10.1103/PhysRevD.104.123023} {\bibfield
  {journal} {\bibinfo  {journal} {Phys. Rev. D}\ }\textbf {\bibinfo {volume}
  {104}},\ \bibinfo {pages} {123023} (\bibinfo {year}
  {2021}{\natexlab{b}})}\BibitemShut {NoStop}%
\bibitem [{\citenamefont {Lacroix}\ \emph {et~al.}(2022)\citenamefont
  {Lacroix}, \citenamefont {Balantekin}, \citenamefont {Cervia}, \citenamefont
  {Patwardhan},\ and\ \citenamefont {Siwach}}]{Lacroix:2022krq}%
  \BibitemOpen
  \bibfield  {author} {\bibinfo {author} {\bibfnamefont {Denis}\ \bibnamefont
  {Lacroix}}, \bibinfo {author} {\bibfnamefont {A.~B.}\ \bibnamefont
  {Balantekin}}, \bibinfo {author} {\bibfnamefont {Michael~J.}\ \bibnamefont
  {Cervia}}, \bibinfo {author} {\bibfnamefont {Amol~V.}\ \bibnamefont
  {Patwardhan}}, \ and\ \bibinfo {author} {\bibfnamefont {Pooja}\ \bibnamefont
  {Siwach}},\ }\bibfield  {title} {\enquote {\bibinfo {title} {{Role of
  non-Gaussian quantum fluctuations in neutrino entanglement}},}\ }\href
  {\doibase 10.1103/PhysRevD.106.123006} {\bibfield  {journal} {\bibinfo
  {journal} {Phys. Rev. D}\ }\textbf {\bibinfo {volume} {106}},\ \bibinfo
  {pages} {123006} (\bibinfo {year} {2022})},\ \Eprint
  {http://arxiv.org/abs/2205.09384} {arXiv:2205.09384 [nucl-th]} \BibitemShut
  {NoStop}%
\bibitem [{\citenamefont {Cervia}\ \emph {et~al.}(2019)\citenamefont {Cervia},
  \citenamefont {Patwardhan}, \citenamefont {Balantekin}, \citenamefont
  {Coppersmith},\ and\ \citenamefont {Johnson}}]{Cervia2019}%
  \BibitemOpen
  \bibfield  {author} {\bibinfo {author} {\bibfnamefont {Michael~J.}\
  \bibnamefont {Cervia}}, \bibinfo {author} {\bibfnamefont {Amol~V.}\
  \bibnamefont {Patwardhan}}, \bibinfo {author} {\bibfnamefont {A.~B.}\
  \bibnamefont {Balantekin}}, \bibinfo {author} {\bibfnamefont {S.~N.}\
  \bibnamefont {Coppersmith}}, \ and\ \bibinfo {author} {\bibfnamefont
  {Calvin~W.}\ \bibnamefont {Johnson}},\ }\bibfield  {title} {\enquote
  {\bibinfo {title} {Entanglement and collective flavor oscillations in a dense
  neutrino gas},}\ }\href {\doibase 10.1103/PhysRevD.100.083001} {\bibfield
  {journal} {\bibinfo  {journal} {Phys. Rev. D}\ }\textbf {\bibinfo {volume}
  {100}},\ \bibinfo {pages} {083001} (\bibinfo {year} {2019})}\BibitemShut
  {NoStop}%
\bibitem [{\citenamefont {Fiorillo}\ and\ \citenamefont
  {Raffelt}(2023)}]{PhysRevD.107.043024}%
  \BibitemOpen
  \bibfield  {author} {\bibinfo {author} {\bibfnamefont {Damiano F.~G.}\
  \bibnamefont {Fiorillo}}\ and\ \bibinfo {author} {\bibfnamefont {Georg~G.}\
  \bibnamefont {Raffelt}},\ }\bibfield  {title} {\enquote {\bibinfo {title}
  {Slow and fast collective neutrino oscillations: Invariants and
  reciprocity},}\ }\href {\doibase 10.1103/PhysRevD.107.043024} {\bibfield
  {journal} {\bibinfo  {journal} {Phys. Rev. D}\ }\textbf {\bibinfo {volume}
  {107}},\ \bibinfo {pages} {043024} (\bibinfo {year} {2023})}\BibitemShut
  {NoStop}%
\bibitem [{\citenamefont {Johns}(2023)}]{Johns:2023ewj}%
  \BibitemOpen
  \bibfield  {author} {\bibinfo {author} {\bibfnamefont {Lucas}\ \bibnamefont
  {Johns}},\ }\bibfield  {title} {\enquote {\bibinfo {title} {{Neutrino
  many-body correlations}},}\ }\href@noop {} {\  (\bibinfo {year} {2023})},\
  \Eprint {http://arxiv.org/abs/2305.04916} {arXiv:2305.04916 [hep-ph]}
  \BibitemShut {NoStop}%
\bibitem [{\citenamefont {Scaramazza}\ \emph {et~al.}(2016)\citenamefont
  {Scaramazza}, \citenamefont {Shastry},\ and\ \citenamefont
  {Yuzbashyan}}]{scaramazza2016integrable}%
  \BibitemOpen
  \bibfield  {author} {\bibinfo {author} {\bibfnamefont {Jasen~A}\ \bibnamefont
  {Scaramazza}}, \bibinfo {author} {\bibfnamefont {B~Sriram}\ \bibnamefont
  {Shastry}}, \ and\ \bibinfo {author} {\bibfnamefont {Emil~A}\ \bibnamefont
  {Yuzbashyan}},\ }\bibfield  {title} {\enquote {\bibinfo {title} {Integrable
  matrix theory: Level statistics},}\ }\href
  {https://journals.aps.org/pre/abstract/10.1103/PhysRevE.94.032106} {\bibfield
   {journal} {\bibinfo  {journal} {Physical Review E}\ }\textbf {\bibinfo
  {volume} {94}},\ \bibinfo {pages} {032106} (\bibinfo {year}
  {2016})}\BibitemShut {NoStop}%
\bibitem [{\citenamefont {Atas}\ \emph {et~al.}(2013)\citenamefont {Atas},
  \citenamefont {Bogomolny}, \citenamefont {Giraud},\ and\ \citenamefont
  {Roux}}]{atas2013distribution}%
  \BibitemOpen
  \bibfield  {author} {\bibinfo {author} {\bibfnamefont {YY}~\bibnamefont
  {Atas}}, \bibinfo {author} {\bibfnamefont {Eugene}\ \bibnamefont
  {Bogomolny}}, \bibinfo {author} {\bibfnamefont {O}~\bibnamefont {Giraud}}, \
  and\ \bibinfo {author} {\bibfnamefont {G}~\bibnamefont {Roux}},\ }\bibfield
  {title} {\enquote {\bibinfo {title} {Distribution of the ratio of consecutive
  level spacings in random matrix ensembles},}\ }\href
  {https://journals.aps.org/prl/abstract/10.1103/PhysRevLett.110.084101}
  {\bibfield  {journal} {\bibinfo  {journal} {Physical review letters}\
  }\textbf {\bibinfo {volume} {110}},\ \bibinfo {pages} {084101} (\bibinfo
  {year} {2013})}\BibitemShut {NoStop}%
\bibitem [{\citenamefont {Deutsch}(1991)}]{PhysRevA.43.2046}%
  \BibitemOpen
  \bibfield  {author} {\bibinfo {author} {\bibfnamefont {J.~M.}\ \bibnamefont
  {Deutsch}},\ }\bibfield  {title} {\enquote {\bibinfo {title} {Quantum
  statistical mechanics in a closed system},}\ }\href {\doibase
  10.1103/PhysRevA.43.2046} {\bibfield  {journal} {\bibinfo  {journal} {Phys.
  Rev. A}\ }\textbf {\bibinfo {volume} {43}},\ \bibinfo {pages} {2046--2049}
  (\bibinfo {year} {1991})}\BibitemShut {NoStop}%
\bibitem [{\citenamefont {Srednicki}(1994)}]{Srednicki1994}%
  \BibitemOpen
  \bibfield  {author} {\bibinfo {author} {\bibfnamefont {Mark}\ \bibnamefont
  {Srednicki}},\ }\bibfield  {title} {\enquote {\bibinfo {title} {Chaos and
  quantum thermalization},}\ }\href {\doibase 10.1103/PhysRevE.50.888}
  {\bibfield  {journal} {\bibinfo  {journal} {Phys. Rev. E}\ }\textbf {\bibinfo
  {volume} {50}},\ \bibinfo {pages} {888--901} (\bibinfo {year}
  {1994})}\BibitemShut {NoStop}%
\bibitem [{\citenamefont {Srednicki}(1996)}]{Srednicki_1996}%
  \BibitemOpen
  \bibfield  {author} {\bibinfo {author} {\bibfnamefont {Mark}\ \bibnamefont
  {Srednicki}},\ }\bibfield  {title} {\enquote {\bibinfo {title} {Thermal
  fluctuations in quantized chaotic systems},}\ }\href {\doibase
  10.1088/0305-4470/29/4/003} {\bibfield  {journal} {\bibinfo  {journal}
  {Journal of Physics A: Mathematical and General}\ }\textbf {\bibinfo {volume}
  {29}},\ \bibinfo {pages} {L75} (\bibinfo {year} {1996})}\BibitemShut
  {NoStop}%
\bibitem [{\citenamefont {D'Alessio}\ \emph {et~al.}(2016)\citenamefont
  {D'Alessio}, \citenamefont {Kafri}, \citenamefont {Polkovnikov},\ and\
  \citenamefont {Rigol}}]{d2016quantum}%
  \BibitemOpen
  \bibfield  {author} {\bibinfo {author} {\bibfnamefont {Luca}\ \bibnamefont
  {D'Alessio}}, \bibinfo {author} {\bibfnamefont {Yariv}\ \bibnamefont
  {Kafri}}, \bibinfo {author} {\bibfnamefont {Anatoli}\ \bibnamefont
  {Polkovnikov}}, \ and\ \bibinfo {author} {\bibfnamefont {Marcos}\
  \bibnamefont {Rigol}},\ }\bibfield  {title} {\enquote {\bibinfo {title} {From
  quantum chaos and eigenstate thermalization to statistical mechanics and
  thermodynamics},}\ }\href@noop {} {\bibfield  {journal} {\bibinfo  {journal}
  {Advances in Physics}\ }\textbf {\bibinfo {volume} {65}},\ \bibinfo {pages}
  {239--362} (\bibinfo {year} {2016})}\BibitemShut {NoStop}%
\bibitem [{\citenamefont {{Rigol}}(2009)}]{2009PhRvA..80e3607R}%
  \BibitemOpen
  \bibfield  {author} {\bibinfo {author} {\bibfnamefont {Marcos}\ \bibnamefont
  {{Rigol}}},\ }\bibfield  {title} {\enquote {\bibinfo {title} {{Quantum
  quenches and thermalization in one-dimensional fermionic systems}},}\ }\href
  {\doibase 10.1103/PhysRevA.80.053607} {\bibfield  {journal} {\bibinfo
  {journal} {\pra}\ }\textbf {\bibinfo {volume} {80}},\ \bibinfo {eid} {053607}
  (\bibinfo {year} {2009})},\ \Eprint {http://arxiv.org/abs/0908.3188}
  {arXiv:0908.3188 [cond-mat.stat-mech]} \BibitemShut {NoStop}%
\bibitem [{\citenamefont {{Rigol}}\ and\ \citenamefont
  {{Santos}}(2010)}]{2010PhRvA..82a1604R}%
  \BibitemOpen
  \bibfield  {author} {\bibinfo {author} {\bibfnamefont {Marcos}\ \bibnamefont
  {{Rigol}}}\ and\ \bibinfo {author} {\bibfnamefont {Lea~F.}\ \bibnamefont
  {{Santos}}},\ }\bibfield  {title} {\enquote {\bibinfo {title} {{Quantum chaos
  and thermalization in gapped systems}},}\ }\href {\doibase
  10.1103/PhysRevA.82.011604} {\bibfield  {journal} {\bibinfo  {journal}
  {\pra}\ }\textbf {\bibinfo {volume} {82}},\ \bibinfo {eid} {011604} (\bibinfo
  {year} {2010})},\ \Eprint {http://arxiv.org/abs/1003.1403} {arXiv:1003.1403
  [cond-mat.stat-mech]} \BibitemShut {NoStop}%
\bibitem [{\citenamefont {{Kim}}\ \emph {et~al.}(2014)\citenamefont {{Kim}},
  \citenamefont {{Ikeda}},\ and\ \citenamefont {{Huse}}}]{2014PhRvE..90e2105K}%
  \BibitemOpen
  \bibfield  {author} {\bibinfo {author} {\bibfnamefont {Hyungwon}\
  \bibnamefont {{Kim}}}, \bibinfo {author} {\bibfnamefont {Tatsuhiko~N.}\
  \bibnamefont {{Ikeda}}}, \ and\ \bibinfo {author} {\bibfnamefont {David~A.}\
  \bibnamefont {{Huse}}},\ }\bibfield  {title} {\enquote {\bibinfo {title}
  {{Testing whether all eigenstates obey the eigenstate thermalization
  hypothesis}},}\ }\href {\doibase 10.1103/PhysRevE.90.052105} {\bibfield
  {journal} {\bibinfo  {journal} {\pre}\ }\textbf {\bibinfo {volume} {90}},\
  \bibinfo {eid} {052105} (\bibinfo {year} {2014})},\ \Eprint
  {http://arxiv.org/abs/1408.0535} {arXiv:1408.0535 [cond-mat.stat-mech]}
  \BibitemShut {NoStop}%
\bibitem [{\citenamefont {{Baldwin}}\ \emph {et~al.}(2017)\citenamefont
  {{Baldwin}}, \citenamefont {{Laumann}}, \citenamefont {{Pal}},\ and\
  \citenamefont {{Scardicchio}}}]{2017PhRvL.118l7201B}%
  \BibitemOpen
  \bibfield  {author} {\bibinfo {author} {\bibfnamefont {C.~L.}\ \bibnamefont
  {{Baldwin}}}, \bibinfo {author} {\bibfnamefont {C.~R.}\ \bibnamefont
  {{Laumann}}}, \bibinfo {author} {\bibfnamefont {A.}~\bibnamefont {{Pal}}}, \
  and\ \bibinfo {author} {\bibfnamefont {A.}~\bibnamefont {{Scardicchio}}},\
  }\bibfield  {title} {\enquote {\bibinfo {title} {{Clustering of Nonergodic
  Eigenstates in Quantum Spin Glasses}},}\ }\href {\doibase
  10.1103/PhysRevLett.118.127201} {\bibfield  {journal} {\bibinfo  {journal}
  {\prl}\ }\textbf {\bibinfo {volume} {118}},\ \bibinfo {eid} {127201}
  (\bibinfo {year} {2017})},\ \Eprint {http://arxiv.org/abs/1611.02296}
  {arXiv:1611.02296 [cond-mat.dis-nn]} \BibitemShut {NoStop}%
\bibitem [{\citenamefont {{{\L}yd{\.Z}ba}}\ \emph {et~al.}(2021)\citenamefont
  {{{\L}yd{\.Z}ba}}, \citenamefont {{Zhang}}, \citenamefont {{Rigol}},\ and\
  \citenamefont {{Vidmar}}}]{2021PhRvB.104u4203L}%
  \BibitemOpen
  \bibfield  {author} {\bibinfo {author} {\bibfnamefont {Patrycja}\
  \bibnamefont {{{\L}yd{\.Z}ba}}}, \bibinfo {author} {\bibfnamefont {Yicheng}\
  \bibnamefont {{Zhang}}}, \bibinfo {author} {\bibfnamefont {Marcos}\
  \bibnamefont {{Rigol}}}, \ and\ \bibinfo {author} {\bibfnamefont {Lev}\
  \bibnamefont {{Vidmar}}},\ }\bibfield  {title} {\enquote {\bibinfo {title}
  {{Single-particle eigenstate thermalization in quantum-chaotic quadratic
  Hamiltonians}},}\ }\href {\doibase 10.1103/PhysRevB.104.214203} {\bibfield
  {journal} {\bibinfo  {journal} {\prb}\ }\textbf {\bibinfo {volume} {104}},\
  \bibinfo {eid} {214203} (\bibinfo {year} {2021})},\ \Eprint
  {http://arxiv.org/abs/2109.06895} {arXiv:2109.06895 [cond-mat.stat-mech]}
  \BibitemShut {NoStop}%
\bibitem [{\citenamefont {{Villase{\~n}or}}\ \emph {et~al.}(2022)\citenamefont
  {{Villase{\~n}or}}, \citenamefont {{Pilatowsky-Cameo}}, \citenamefont
  {{Bastarrachea-Magnani}}, \citenamefont {{Lerma-Hern{\'a}ndez}},
  \citenamefont {{Santos}},\ and\ \citenamefont
  {{Hirsch}}}]{2022Entrp..25....8V}%
  \BibitemOpen
  \bibfield  {author} {\bibinfo {author} {\bibfnamefont {David}\ \bibnamefont
  {{Villase{\~n}or}}}, \bibinfo {author} {\bibfnamefont {Sa{\'u}l}\
  \bibnamefont {{Pilatowsky-Cameo}}}, \bibinfo {author} {\bibfnamefont
  {Miguel~A.}\ \bibnamefont {{Bastarrachea-Magnani}}}, \bibinfo {author}
  {\bibfnamefont {Sergio}\ \bibnamefont {{Lerma-Hern{\'a}ndez}}}, \bibinfo
  {author} {\bibfnamefont {Lea~F.}\ \bibnamefont {{Santos}}}, \ and\ \bibinfo
  {author} {\bibfnamefont {Jorge~G.}\ \bibnamefont {{Hirsch}}},\ }\bibfield
  {title} {\enquote {\bibinfo {title} {{Chaos and Thermalization in the
  Spin-Boson Dicke Model}},}\ }\href {\doibase 10.3390/e25010008} {\bibfield
  {journal} {\bibinfo  {journal} {Entropy}\ }\textbf {\bibinfo {volume} {25}},\
  \bibinfo {pages} {8} (\bibinfo {year} {2022})},\ \Eprint
  {http://arxiv.org/abs/2211.08434} {arXiv:2211.08434 [quant-ph]} \BibitemShut
  {NoStop}%
\bibitem [{\citenamefont {Shiraishi}\ and\ \citenamefont
  {Mori}(2017)}]{shiraishi2017systematic}%
  \BibitemOpen
  \bibfield  {author} {\bibinfo {author} {\bibfnamefont {Naoto}\ \bibnamefont
  {Shiraishi}}\ and\ \bibinfo {author} {\bibfnamefont {Takashi}\ \bibnamefont
  {Mori}},\ }\bibfield  {title} {\enquote {\bibinfo {title} {Systematic
  construction of counterexamples to the eigenstate thermalization
  hypothesis},}\ }\href@noop {} {\bibfield  {journal} {\bibinfo  {journal}
  {Physical review letters}\ }\textbf {\bibinfo {volume} {119}},\ \bibinfo
  {pages} {030601} (\bibinfo {year} {2017})}\BibitemShut {NoStop}%
\bibitem [{\citenamefont {Pakrouski}\ \emph {et~al.}(2020)\citenamefont
  {Pakrouski}, \citenamefont {Pallegar}, \citenamefont {Popov},\ and\
  \citenamefont {Klebanov}}]{pakrouski2020many}%
  \BibitemOpen
  \bibfield  {author} {\bibinfo {author} {\bibfnamefont {Kiryl}\ \bibnamefont
  {Pakrouski}}, \bibinfo {author} {\bibfnamefont {Preethi~N}\ \bibnamefont
  {Pallegar}}, \bibinfo {author} {\bibfnamefont {Fedor~K}\ \bibnamefont
  {Popov}}, \ and\ \bibinfo {author} {\bibfnamefont {Igor~R}\ \bibnamefont
  {Klebanov}},\ }\bibfield  {title} {\enquote {\bibinfo {title} {Many-body
  scars as a group invariant sector of hilbert space},}\ }\href@noop {}
  {\bibfield  {journal} {\bibinfo  {journal} {Physical review letters}\
  }\textbf {\bibinfo {volume} {125}},\ \bibinfo {pages} {230602} (\bibinfo
  {year} {2020})}\BibitemShut {NoStop}%
\bibitem [{\citenamefont {Regnault}\ \emph {et~al.}(2022)\citenamefont
  {Regnault}, \citenamefont {Moudgalya},\ and\ \citenamefont
  {Bernevig}}]{regnault2022quantum}%
  \BibitemOpen
  \bibfield  {author} {\bibinfo {author} {\bibfnamefont {Nicolas}\ \bibnamefont
  {Regnault}}, \bibinfo {author} {\bibfnamefont {Sanjay}\ \bibnamefont
  {Moudgalya}}, \ and\ \bibinfo {author} {\bibfnamefont {B~Andrei}\
  \bibnamefont {Bernevig}},\ }\bibfield  {title} {\enquote {\bibinfo {title}
  {Quantum many-body scars and hilbert space fragmentation: a review of exact
  results},}\ }\href@noop {} {\bibfield  {journal} {\bibinfo  {journal}
  {Reports on Progress in Physics}\ } (\bibinfo {year} {2022})}\BibitemShut
  {NoStop}%
\bibitem [{\citenamefont {Srednicki}(1999)}]{srednicki1999approach}%
  \BibitemOpen
  \bibfield  {author} {\bibinfo {author} {\bibfnamefont {Mark}\ \bibnamefont
  {Srednicki}},\ }\bibfield  {title} {\enquote {\bibinfo {title} {The approach
  to thermal equilibrium in quantized chaotic systems},}\ }\href
  {https://iopscience.iop.org/article/10.1088/0305-4470/32/7/007/meta}
  {\bibfield  {journal} {\bibinfo  {journal} {Journal of Physics A:
  Mathematical and General}\ }\textbf {\bibinfo {volume} {32}},\ \bibinfo
  {pages} {1163} (\bibinfo {year} {1999})}\BibitemShut {NoStop}%
\bibitem [{\citenamefont {{Parker}}\ \emph {et~al.}(2019)\citenamefont
  {{Parker}}, \citenamefont {{Cao}}, \citenamefont {{Avdoshkin}}, \citenamefont
  {{Scaffidi}},\ and\ \citenamefont {{Altman}}}]{2019PhRvX9d1017P}%
  \BibitemOpen
  \bibfield  {author} {\bibinfo {author} {\bibfnamefont {Daniel~E.}\
  \bibnamefont {{Parker}}}, \bibinfo {author} {\bibfnamefont {Xiangyu}\
  \bibnamefont {{Cao}}}, \bibinfo {author} {\bibfnamefont {Alexander}\
  \bibnamefont {{Avdoshkin}}}, \bibinfo {author} {\bibfnamefont {Thomas}\
  \bibnamefont {{Scaffidi}}}, \ and\ \bibinfo {author} {\bibfnamefont {Ehud}\
  \bibnamefont {{Altman}}},\ }\bibfield  {title} {\enquote {\bibinfo {title}
  {{A Universal Operator Growth Hypothesis}},}\ }\href {\doibase
  10.1103/PhysRevX.9.041017} {\bibfield  {journal} {\bibinfo  {journal}
  {Physical Review X}\ }\textbf {\bibinfo {volume} {9}},\ \bibinfo {eid}
  {041017} (\bibinfo {year} {2019})},\ \Eprint
  {http://arxiv.org/abs/1812.08657} {arXiv:1812.08657 [cond-mat.stat-mech]}
  \BibitemShut {NoStop}%
\bibitem [{\citenamefont {Murthy}\ and\ \citenamefont
  {Srednicki}(2019)}]{PhysRevLett.123.230606}%
  \BibitemOpen
  \bibfield  {author} {\bibinfo {author} {\bibfnamefont {Chaitanya}\
  \bibnamefont {Murthy}}\ and\ \bibinfo {author} {\bibfnamefont {Mark}\
  \bibnamefont {Srednicki}},\ }\bibfield  {title} {\enquote {\bibinfo {title}
  {Bounds on chaos from the eigenstate thermalization hypothesis},}\ }\href
  {\doibase 10.1103/PhysRevLett.123.230606} {\bibfield  {journal} {\bibinfo
  {journal} {Phys. Rev. Lett.}\ }\textbf {\bibinfo {volume} {123}},\ \bibinfo
  {pages} {230606} (\bibinfo {year} {2019})}\BibitemShut {NoStop}%
\bibitem [{\citenamefont {Khatami}\ \emph {et~al.}(2013)\citenamefont
  {Khatami}, \citenamefont {Pupillo}, \citenamefont {Srednicki},\ and\
  \citenamefont {Rigol}}]{PhysRevLett.111.050403}%
  \BibitemOpen
  \bibfield  {author} {\bibinfo {author} {\bibfnamefont {Ehsan}\ \bibnamefont
  {Khatami}}, \bibinfo {author} {\bibfnamefont {Guido}\ \bibnamefont
  {Pupillo}}, \bibinfo {author} {\bibfnamefont {Mark}\ \bibnamefont
  {Srednicki}}, \ and\ \bibinfo {author} {\bibfnamefont {Marcos}\ \bibnamefont
  {Rigol}},\ }\bibfield  {title} {\enquote {\bibinfo {title}
  {Fluctuation-dissipation theorem in an isolated system of quantum dipolar
  bosons after a quench},}\ }\href {\doibase 10.1103/PhysRevLett.111.050403}
  {\bibfield  {journal} {\bibinfo  {journal} {Phys. Rev. Lett.}\ }\textbf
  {\bibinfo {volume} {111}},\ \bibinfo {pages} {050403} (\bibinfo {year}
  {2013})}\BibitemShut {NoStop}%
\bibitem [{\citenamefont {Murthy}\ \emph {et~al.}(2023)\citenamefont {Murthy},
  \citenamefont {Babakhani}, \citenamefont {Iniguez}, \citenamefont
  {Srednicki},\ and\ \citenamefont {Yunger~Halpern}}]{Murthy:2022dao}%
  \BibitemOpen
  \bibfield  {author} {\bibinfo {author} {\bibfnamefont {Chaitanya}\
  \bibnamefont {Murthy}}, \bibinfo {author} {\bibfnamefont {Arman}\
  \bibnamefont {Babakhani}}, \bibinfo {author} {\bibfnamefont {Fernando}\
  \bibnamefont {Iniguez}}, \bibinfo {author} {\bibfnamefont {Mark}\
  \bibnamefont {Srednicki}}, \ and\ \bibinfo {author} {\bibfnamefont {Nicole}\
  \bibnamefont {Yunger~Halpern}},\ }\bibfield  {title} {\enquote {\bibinfo
  {title} {Non-abelian eigenstate thermalization hypothesis},}\ }\href
  {\doibase 10.1103/PhysRevLett.130.140402} {\bibfield  {journal} {\bibinfo
  {journal} {Phys. Rev. Lett.}\ }\textbf {\bibinfo {volume} {130}},\ \bibinfo
  {pages} {140402} (\bibinfo {year} {2023})}\BibitemShut {NoStop}%
\bibitem [{\citenamefont {Giovannetti}\ \emph {et~al.}(2003)\citenamefont
  {Giovannetti}, \citenamefont {Lloyd},\ and\ \citenamefont
  {Maccone}}]{Giovannetti_2003b}%
  \BibitemOpen
  \bibfield  {author} {\bibinfo {author} {\bibfnamefont {Vittorio}\
  \bibnamefont {Giovannetti}}, \bibinfo {author} {\bibfnamefont {Seth}\
  \bibnamefont {Lloyd}}, \ and\ \bibinfo {author} {\bibfnamefont {Lorenzo}\
  \bibnamefont {Maccone}},\ }\bibfield  {title} {\enquote {\bibinfo {title}
  {Quantum limits to dynamical evolution},}\ }\href
  {https://journals.aps.org/pra/abstract/10.1103/PhysRevA.67.052109} {\bibfield
   {journal} {\bibinfo  {journal} {Physical Review A}\ }\textbf {\bibinfo
  {volume} {67}} (\bibinfo {year} {2003})}\BibitemShut {NoStop}%
\bibitem [{\citenamefont {Guti\'errez}\ and\ \citenamefont
  {Goussev}(2009)}]{PhysRevE.79.046211}%
  \BibitemOpen
  \bibfield  {author} {\bibinfo {author} {\bibfnamefont {Martha}\ \bibnamefont
  {Guti\'errez}}\ and\ \bibinfo {author} {\bibfnamefont {Arseni}\ \bibnamefont
  {Goussev}},\ }\bibfield  {title} {\enquote {\bibinfo {title} {Long-time
  saturation of the loschmidt echo in quantum chaotic billiards},}\ }\href
  {\doibase 10.1103/PhysRevE.79.046211} {\bibfield  {journal} {\bibinfo
  {journal} {Phys. Rev. E}\ }\textbf {\bibinfo {volume} {79}},\ \bibinfo
  {pages} {046211} (\bibinfo {year} {2009})}\BibitemShut {NoStop}%
\bibitem [{\citenamefont {Houcke}\ \emph {et~al.}(2010)\citenamefont {Houcke},
  \citenamefont {Kozik}, \citenamefont {Prokof'ev},\ and\ \citenamefont
  {Svistunov}}]{VanHoucke:2010}%
  \BibitemOpen
  \bibfield  {author} {\bibinfo {author} {\bibfnamefont {Kris~Van}\
  \bibnamefont {Houcke}}, \bibinfo {author} {\bibfnamefont {Evgeny}\
  \bibnamefont {Kozik}}, \bibinfo {author} {\bibfnamefont {N.}~\bibnamefont
  {Prokof'ev}}, \ and\ \bibinfo {author} {\bibfnamefont {B.}~\bibnamefont
  {Svistunov}},\ }\bibfield  {title} {\enquote {\bibinfo {title} {Diagrammatic
  monte carlo},}\ }\href {\doibase 10.1016/j.phpro.2010.09.034} {\bibfield
  {journal} {\bibinfo  {journal} {Physics Procedia}\ }\textbf {\bibinfo
  {volume} {6}},\ \bibinfo {pages} {95--105} (\bibinfo {year}
  {2010})}\BibitemShut {NoStop}%
\bibitem [{\citenamefont {Martin}\ \emph {et~al.}(2023)\citenamefont {Martin},
  \citenamefont {Roggero}, \citenamefont {Duan},\ and\ \citenamefont
  {Carlson}}]{Martin:2023ljq}%
  \BibitemOpen
  \bibfield  {author} {\bibinfo {author} {\bibfnamefont {Joshua~D.}\
  \bibnamefont {Martin}}, \bibinfo {author} {\bibfnamefont {A.}~\bibnamefont
  {Roggero}}, \bibinfo {author} {\bibfnamefont {Huaiyu}\ \bibnamefont {Duan}},
  \ and\ \bibinfo {author} {\bibfnamefont {J.}~\bibnamefont {Carlson}},\
  }\bibfield  {title} {\enquote {\bibinfo {title} {{Many-body neutrino flavor
  entanglement in a simple dynamic model}},}\ }\href@noop {} {\  (\bibinfo
  {year} {2023})},\ \Eprint {http://arxiv.org/abs/2301.07049} {arXiv:2301.07049
  [hep-ph]} \BibitemShut {NoStop}%
\bibitem [{\citenamefont {Duan}\ \emph
  {et~al.}(2006{\natexlab{b}})\citenamefont {Duan}, \citenamefont {Fuller},\
  and\ \citenamefont {Qian}}]{Duan:2005cp}%
  \BibitemOpen
  \bibfield  {author} {\bibinfo {author} {\bibfnamefont {Huaiyu}\ \bibnamefont
  {Duan}}, \bibinfo {author} {\bibfnamefont {George~M.}\ \bibnamefont
  {Fuller}}, \ and\ \bibinfo {author} {\bibfnamefont {Yong-Zhong}\ \bibnamefont
  {Qian}},\ }\bibfield  {title} {\enquote {\bibinfo {title} {{Collective
  neutrino flavor transformation in supernovae}},}\ }\href {\doibase
  10.1103/PhysRevD.74.123004} {\bibfield  {journal} {\bibinfo  {journal} {Phys.
  Rev. D}\ }\textbf {\bibinfo {volume} {74}},\ \bibinfo {pages} {123004}
  (\bibinfo {year} {2006}{\natexlab{b}})},\ \Eprint
  {http://arxiv.org/abs/astro-ph/0511275} {arXiv:astro-ph/0511275} \BibitemShut
  {NoStop}%
\bibitem [{\citenamefont {Hall}\ \emph {et~al.}(2021)\citenamefont {Hall},
  \citenamefont {Roggero}, \citenamefont {Baroni},\ and\ \citenamefont
  {Carlson}}]{Hall2021}%
  \BibitemOpen
  \bibfield  {author} {\bibinfo {author} {\bibfnamefont {Benjamin}\
  \bibnamefont {Hall}}, \bibinfo {author} {\bibfnamefont {Alessandro}\
  \bibnamefont {Roggero}}, \bibinfo {author} {\bibfnamefont {Alessandro}\
  \bibnamefont {Baroni}}, \ and\ \bibinfo {author} {\bibfnamefont {Joseph}\
  \bibnamefont {Carlson}},\ }\bibfield  {title} {\enquote {\bibinfo {title}
  {Simulation of collective neutrino oscillations on a quantum computer},}\
  }\href {\doibase 10.1103/PhysRevD.104.063009} {\bibfield  {journal} {\bibinfo
   {journal} {Phys. Rev. D}\ }\textbf {\bibinfo {volume} {104}},\ \bibinfo
  {pages} {063009} (\bibinfo {year} {2021})}\BibitemShut {NoStop}%
\bibitem [{\citenamefont {Yeter-Aydeniz}\ \emph {et~al.}(2022)\citenamefont
  {Yeter-Aydeniz}, \citenamefont {Bangar}, \citenamefont {Siopsis},\ and\
  \citenamefont {C~Pooser}}]{yeter2022collective}%
  \BibitemOpen
  \bibfield  {author} {\bibinfo {author} {\bibfnamefont {K{\"u}bra}\
  \bibnamefont {Yeter-Aydeniz}}, \bibinfo {author} {\bibfnamefont {Shikha}\
  \bibnamefont {Bangar}}, \bibinfo {author} {\bibfnamefont {George}\
  \bibnamefont {Siopsis}}, \ and\ \bibinfo {author} {\bibfnamefont {Raphael}\
  \bibnamefont {C~Pooser}},\ }\bibfield  {title} {\enquote {\bibinfo {title}
  {Collective neutrino oscillations on a quantum computer},}\ }\href {\doibase
  10.1007/s11128-021-03348-x} {\bibfield  {journal} {\bibinfo  {journal}
  {Quantum Information Processing}\ }\textbf {\bibinfo {volume} {21}},\
  \bibinfo {pages} {84} (\bibinfo {year} {2022})}\BibitemShut {NoStop}%
\bibitem [{\citenamefont {Illa}\ and\ \citenamefont
  {Savage}(2022)}]{illa2022basic}%
  \BibitemOpen
  \bibfield  {author} {\bibinfo {author} {\bibfnamefont {Marc}\ \bibnamefont
  {Illa}}\ and\ \bibinfo {author} {\bibfnamefont {Martin~J.}\ \bibnamefont
  {Savage}},\ }\bibfield  {title} {\enquote {\bibinfo {title} {Basic elements
  for simulations of standard-model physics with quantum annealers: Multigrid
  and clock states},}\ }\href {\doibase 10.1103/PhysRevA.106.052605} {\bibfield
   {journal} {\bibinfo  {journal} {Phys. Rev. A}\ }\textbf {\bibinfo {volume}
  {106}},\ \bibinfo {pages} {052605} (\bibinfo {year} {2022})}\BibitemShut
  {NoStop}%
\bibitem [{\citenamefont {Amitrano}\ \emph {et~al.}(2023)\citenamefont
  {Amitrano}, \citenamefont {Roggero}, \citenamefont {Luchi}, \citenamefont
  {Turro}, \citenamefont {Vespucci},\ and\ \citenamefont
  {Pederiva}}]{Amitrano2023}%
  \BibitemOpen
  \bibfield  {author} {\bibinfo {author} {\bibfnamefont {Valentina}\
  \bibnamefont {Amitrano}}, \bibinfo {author} {\bibfnamefont {Alessandro}\
  \bibnamefont {Roggero}}, \bibinfo {author} {\bibfnamefont {Piero}\
  \bibnamefont {Luchi}}, \bibinfo {author} {\bibfnamefont {Francesco}\
  \bibnamefont {Turro}}, \bibinfo {author} {\bibfnamefont {Luca}\ \bibnamefont
  {Vespucci}}, \ and\ \bibinfo {author} {\bibfnamefont {Francesco}\
  \bibnamefont {Pederiva}},\ }\bibfield  {title} {\enquote {\bibinfo {title}
  {Trapped-ion quantum simulation of collective neutrino oscillations},}\
  }\href {\doibase 10.1103/PhysRevD.107.023007} {\bibfield  {journal} {\bibinfo
   {journal} {Phys. Rev. D}\ }\textbf {\bibinfo {volume} {107}},\ \bibinfo
  {pages} {023007} (\bibinfo {year} {2023})}\BibitemShut {NoStop}%
\bibitem [{\citenamefont {Siwach}\ \emph {et~al.}(2023)\citenamefont {Siwach},
  \citenamefont {Harrison},\ and\ \citenamefont
  {Balantekin}}]{siwach2023collective}%
  \BibitemOpen
  \bibfield  {author} {\bibinfo {author} {\bibfnamefont {Pooja}\ \bibnamefont
  {Siwach}}, \bibinfo {author} {\bibfnamefont {Kaytlin}\ \bibnamefont
  {Harrison}}, \ and\ \bibinfo {author} {\bibfnamefont {A.~Baha}\ \bibnamefont
  {Balantekin}},\ }\href@noop {} {\enquote {\bibinfo {title} {Collective
  neutrino oscillations on a quantum computer with hybrid quantum-classical
  algorithm},}\ } (\bibinfo {year} {2023}),\ \Eprint
  {http://arxiv.org/abs/2308.09123} {arXiv:2308.09123 [quant-ph]} \BibitemShut
  {NoStop}%
\end{thebibliography}%
\begin{widetext}
    
\appendix
\section{Short Time Evolution}\label{app:time}
An important question to address is when equilibration is reached for a given initial state. A simple approach is to Taylor expand the time evolution of the operator in the given initial state, and use the expansion to estimate when  the first crossing of the equilibrium value is reached, as depicted in Fig.~\ref{fig:sigz_v_t}. In what follows, we will refer to the flavor state of a given neutrino simply as the state of a given site in our system. We introduce the swap operator $\swap{ij}= \frac{1}{2}\hat{\sigv}_{i} \cdot \hat{\sigv}_{j}+\frac{\mathbb{1}}{2} $, whose action on a tensor product state of sites $i$ \& $j$ and any other state representing the rest of the system is given by:
\begin{align}
\swap{ij}|a\rangle_{i}\otimes|b\rangle_{j}\otimes|\psi\rangle=|b\rangle_{i}\otimes|a\rangle_{j}\otimes|\psi\rangle\,.
\end{align}
We can take our Hamiltonian of Eq.~\eqref{eq:Hvv} and rewrite it as (up to a term proportional to the identity):
\begin{equation} \label{eq:Hvv_swap}
    \Hvv = \frac{1}{N} \sum_{i<j}^{N} \mu_{ij} \swap{ij} \, ,
\end{equation}
where we introduce the variables $\mu_{ij} = \mu(1-\vvec_{i}\cdot\vvec_{j})$ for conciseness.

Working in the Heisenberg picture, we then can calculate the time evolution of a Pauli-matrix at a site $i$, which we fix for definiteness to be in the $\ez$-direction and choose $i=1$,
though our results do not depend on these choices:
\begin{align}
\hat{\sigma}_{3,1}(t)&=e^{-\I \Hvv t}\hat{\sigma}_{3,1} e^{\I \Hvv t}=\hat{\sigma}_{3,1} - \I t[\Hvv,\hat{\sigma}_{3,1}]+\frac{(\I t)^2}{2}[\Hvv,[\Hvv,\hat{\sigma}_{3,1}]]+...\,\,.
\end{align}
To compute these commutators, we note the following identity:
\begin{align}\label{eq:commutate_swaps}
  [ \Hvv,\hat{P}\cdot\hat{\sigma}_{3,k} ]&=\sum_{i<j}^{N}\frac{\mu_{ij}}{N}\Big(\swap{ij} \hat{P} \hat{\sigma}_{3,k} - \hat{P} \swap{ij} \hat{\sigma}_{3, \rho_{ij}(k)}\Big)\\
  \textrm{with index: } \rho_{ij}(k)&=\begin{cases}
    j\text{ if } k=i,\\
    i\text{ if } k=j,\\
    k\text{ else } 
  \end{cases} .
\end{align}  
Here, $\hat{P}$ is an extended permutation operator built from a product of multiple swaps, and we have shifted our Pauli matrix to operate on another site using:
\begin{align}
\hat{\sigma}_{3,k} \swap{ij}&=\swap{ij} \hat{\sigma}_{3,\rho_{ij}(k)}\,.
\end{align}
This allows us to express the commutators as a sum over permutations times a Pauli matrix. We then compute:
\begin{align}\label{eq:commutate_swaps_linear}
  [\Hvv,\hat{\sigma}_{3,1}]=&\sum_{1<j\leq N}\frac{\mu_{1j}}{N}\swap{1j}\Big(\hat{\sigma}_{3,1}-\hat{\sigma}_{3,j}\Big)\,,\\
\label{eq:commutate_swaps_quad}
[\Hvv,[\Hvv,\hat{\sigma}_{3,1}]] =&\frac{1}{N^{2}}\!\! \sum_{1<j\leq N}\sum_{\ell\neq 1,j}\! \Big\{ \mu_{1j}\mu_{j\ell}
    \big(\swap{1j\ell}\left(\hat{\sigma}_{3,1}-\hat{\sigma}_{3,j}\right)\!-\!\swap{1\ell j}\left(\hat{\sigma}_{3,1}-\hat{\sigma}_{3,\ell}\right)\big)\nonumber\\
    &\qquad\qquad\qquad+  \mu_{1j}\mu_{1\ell}  
    \big(\swap{1j\ell}\left(\hat{\sigma}_{3,1}-\hat{\sigma}_{3,j}\right)\!-\!\swap{1\ell j}\left(\hat{\sigma}_{3,j}-\hat{\sigma}_{3,\ell}\right)\big) \Big\}+2\!\!\!\!\sum_{1<j\leq N}\!\frac{\mu_{1j}^2}{N^2}\big(\hat{\sigma}_{3,1}-\hat{\sigma}_{3,j}\big).
\end{align}

The operators of the form $\swap{ijk} \equiv \swap{ik} \swap{ij}$ are cyclic permutations, mapping states as from site $i$ to $j$, site $j$ to $k$, and site $k$ to $i$.

Using the operator norm (the largest absolute value of the eigenvalues in the operator, $\norm{\hat{O}}=\text{sup}_{\psi}|\langle\psi|\hat{O}|\psi\rangle|$), we can bound how large the expectation values for any given state using the triangle inequality and sub-multiplicativity of the norm:
{\small\begin{align}
\normx{\hat{A}+\hat{B}}\leq \normx{\hat{A}}+\normx{\hat{B}}\,,\\
\normx{\hat{A} \hat{B}}\leq \normx{\hat{A}} \normx{\hat{B}}\,,
\end{align}}
to get:

{\small\begin{align}
\normx{[\Hvv,\hat{\sigma}_{3,1}]}&\leq 2\sum_{1<j\leq N}\frac{\mu_{1j}}{N}\sim \mathcal{O}(\mu)\,,\\
\normx{[\Hvv,[\Hvv,\hat{\sigma}_{3,1}]]} &\leq 
\frac{4}{N^2} \sum_{1<j\leq N} \Big\{  \mu_{1j}^2  
     + \sum_{\ell\neq 1,j} 
    \big( \mu_{1j}\mu_{j\ell} + \mu_{1j}\mu_{1\ell} \big) \Big\} 
    \sim \mathcal{O}(\mu^2)\,.
\end{align}}

These represent \emph{worst/best} case scenario for the expectation value of these operators, that is, the largest values the coefficients of the Taylor expansion can take.
However, a given state may or may not have an expectation value with the same asymptotic behavior in the limit that $N\gg 1$ as our upper bounds. To see this, we consider the product state with $\pm 1$ states at all sites, polarized along the 
$\ez$-axis:
\begin{align}\label{eq:specific_states}
|m_{+},m_{-}\rangle = |\!\!\underbrace{1,...,1}_{m_{+}\text{ times }},\underbrace{-1,...,-1}_{m_{-}\text{ times }}\rangle\,.
\end{align}  
There is nothing special about this product state in this basis; any permutation of the sites states will lead to the same conclusions in what follows. We can quickly see:
\begin{align}
 \langle m_{+},m_{-}|[\Hvv,\hat{\sigma}_{3,1}]|m_{+},m_{-}\rangle = 0\,.
\end{align}
The swap operator forces the states at sites $1$ and $j$ to be identical, but then the expectation values of the Pauli-matricies at those sites are also identical. Similarly, we compute:
\begin{align}
 \langle m_{+},m_{-}|&[\Hvv,[\Hvv,\hat{\sigma}_{3,1}]]|m_{+},m_{-}\rangle= 4\sum_{j, |\phi\rangle_j=|-1\rangle}\frac{\mu_{1j}^2}{N^2}\sim \mathcal{O}\Big(\frac{\mu^2 m_{-}}{N^2}\Big)\,.
\end{align}
The contribution from the first two lines of Eq.~\eqref{eq:commutate_swaps_quad} is zero, term by term in the sums. The permutation operators force the states to be identical at the permuting sites, but then the expectation value of the Pauli matrices must cancel.

In general, moving beyond a single flavor polarized product state, we can extend these conclusions to the class of states that have random phases when written in the computational basis of product states of $|\pm 1\rangle$. The expectation value becomes approximately diagonal in the product states, with positive coefficients bounded by $1$ from unitarity, so for such states, we can expect:
{\small\begin{align}
    \langle\psi|[\Hvv,\hat{\sigma}_{3,1}]|\psi\rangle&\sim 0\,, \textrm{ and}\\
    \langle\psi|[\Hvv,[\Hvv,\hat{\sigma}_{3,1}]]|\psi\rangle&\sim \mathcal{O}\Big(\frac{\mu^2}{N}\Big)\,.
    \end{align}}
In the second line, we assume that the product states generically have $\mathcal{O}(N)$ $\ket{+1}$-polarized states and/or $\mathcal{O}(N)$ $\ket{-1}$-polarized states. Clearly they cannot exceed this.
\end{widetext}

\end{document}